%% file: Forte_pp.tex
\documentclass[preprint2,titlepage]{aastex}
\usepackage{graphicx}
\shortauthors{Forte, Bassino, Vega, Pellizza, Cellone, M\'endez}
\shorttitle{Polarimetric Survey of 47 Tucanae}

\begin{document}

\title{A Polarimetric Survey for Dust in 47 Tucanae (NGC
104)\footnote{Based on observations made at the Complejo Astron\'omico
El Leoncito, which is operated under agreement between CONICET and the
National Universities of La Plata, C\'ordoba, and San Juan.}}

\author{ Juan C. Forte }
\affil{Facultad de Cs.\ Astron\'omicas y Geof\'{\i}sicas, UNLP,
and CONICET, Argentina\\
e-mail: forte@fcaglp.unlp.edu.ar}

\author{ Lilia P. Bassino }
\affil{Facultad de Cs.\ Astron\'omicas y Geof\'{\i}sicas, UNLP,
and IALP-CONICET, Argentina%\\
%e-mail: lbassino@fcaglp.unlp.edu.ar
}

\author{ E. Irene Vega }
\affil{ Instituto de Astronom\'{\i}a y F\'{\i}sica del Espacio, CONICET,
and Facultad de Cs.\ Astron\'omicas y Geof\'{\i}sicas, UNLP,
Argentina%\\
%e-mail: irene@iafe.uba.ar
 }

\author{ Leonardo J. Pellizza Gonz\'alez }
\affil{ Instituto de Astronom\'{\i}a y F\'{\i}sica del Espacio, CONICET,
Argentina%\\
%e-mail: pellizza@iafe.uba.ar
 }

\author{ Sergio A. Cellone }
\affil{ Facultad de Cs.\ Astron\'omicas y Geof\'{\i}sicas, UNLP,
and IALP-CONICET, Argentina%\\
%e-mail: scellone@fcaglp.unlp.edu.ar
 }

\author{ Mariano M\'endez }
\affil{ SRON National Institute for Space Research, Sorbonnelaan 2, NL-3584 CA,
Utrecht, the Netherlands%\\
%e-mail: M.Mendez@sron.nl
}

\begin{abstract}
We present linear polarization in the $V$ band for 77 stars in
the field of the globular cluster 47\,Tucanae (NGC 104), and
 for 14 bright-star-free regions, located along an elliptical
isophotal contour of the cluster, as well as $\ubvr I$ measurements for the
cluster nucleus. The observations show variable foreground
polarization that, once removed, leaves marginally significant
polarization residuals for the non-variable bright red giants.
Although these residuals are small there is, however, a systematic
trend in the sense that the larger ones are seen towards the south of the
cluster (in a direction opposite to that of the cluster proper motion).
In contrast, most of the variable stars do show significant intrinsic
polarization. The behavior of the star-free regions is similar to that
of the non-variable stars and sets an upper limit to the possible
existence of a global pattern of scattered (and polarized) intra-cluster
light in the $V$ band. In turn, the multicolor observations of the cluster
nucleus cannot be fitted with a Serkowski law and exhibit a polarization
excess both in the $U$ and $B$ bands. This polarization could be explained
as a combination of the foreground interstellar component and another one
arising in dust located in the nucleus and illuminated by a bright blue
post-asymptotic star (at $48\arcsec$ from the cluster center). An
inspection of a set of archive HST WFPC2 images reveals the presence
of a number of dark patches in the innermost regions of the cluster.
A prominent patch (some $5\arcsec$ by $3\arcsec$ in size) located
at $12\arcsec$ from the cluster center and with a position angle (N to E)
of $120^\circ$, has a slightly different polarization, compared to that of
the cluster nucleus, and appears as a  good candidate to be identified as
a dust globule within the cluster.

\end{abstract}


\section{Introduction \label{s_Intro}}

Arguments in favor and against the presence of dust within  globular
clusters have been given along the years. Although an increasing volume
of recent infrared observations suggests that the dust content should
be very small, if any, the number of hypothesis regarding the nature
of the dust particles (presumably originated by mass loss processes
in luminous red giants), their spatial distribution, equilibrium
temperatures, etc., render these dust mass estimates as still
 uncertain.
A good example of this situation is depicted in the classical paper
by \citet{G88} who discuss IRAS observations of the globular cluster
47\,Tucanae (NGC 104) and report the existence of excess infrared
emission at $100~ \micron$ attributed to dust thermal emission. That
result can be confronted with more recent ISO observations that reach
 $120~\micron$ \citep{ISO}, and do not confirm the presence of such excess.
 Both papers are illustrative of the difficulties,
 among others,
 involved in the subtraction of the stellar background \citep*[see also][]
{K95}.

Among the arguments in favor of dust within globulars, and starting
with the statistical approach by \citet{ROB}, we can mention the
multicolor photometric observations by \citet{KW78} and \citet{FM88}, or
the polarimetry by \citet{MS81} and \citet{FM89}. In turn, \citet{AL90},
disputed these results and pointed out that, in some cases, neither
the photometric nor the polarimetric observations could yield
an unambiguous answer to the nature of the dark patches seen in
several globular clusters, i.e., if they are real dust clouds or just
statistical fluctuations of the stellar distribution or, alternatively,
if the dust is inside  or simply along the line of sight to the cluster.

This paper presents a linear polarization survey of the
globular cluster 47\,Tucanae (NGC 104). Because of its high metallicity
and high mass, this cluster appears as a good candidate
to form and retain (at least some) dust. The structure of the survey
includes three different components, namely,

\begin{enumerate}

\item     The search for intrinsic polarization in the most luminous red
     giants that, due to mass loss processes, could form and eject
     dust particles. The integrated scattered light will be polarized
     only if the resulting dust shells from those processes are not
     spherically symmetric around the star.
     This kind of survey has been carried out (with negative
     results) by \citet*{MI90} and \citet*{MI92} and, in the case
     of 47\,Tuc itself, by \citet{OR97} who, however, mention
     the possible detection of intrinsic polarization in some of its
     red variable stars.

\item     The search for tangential polarization (i.e., perpendicular
     to the direction to the cluster center at a given position) arising
     in dust illuminated by the overall radiation field. The use of this
     effect as a diagnosis tool for dust within stellar systems
     was proposed originally by \citet{JU78}. For example, the
     existence of low optical depth dust distributed around the
     cluster center (and following some kind of radial dependence)
     would produce a regular pattern of the polarization $P$, i.e., a
     double wave behavior of the $U$ and $Q$ Stokes parameters
     defined as $U=P\sin(2\theta)$ and $Q=P\cos(2\theta)$,
     with position angle on the sky. For this purpose
     a number of bright-star-free regions
     were observed following, approximately, an isophotal contour of the
     cluster and avoiding bright stars (no star brighter than
     $V \sim 15.0$ could be seen within the diaphragm).

\item     A multicolor analysis of the cluster nucleus using several
     aperture diaphragms ranging from $17\arcsec$ to $45\arcsec$
     in diameter. A wavelength dependence of the
     polarization should show the feature of interstellar
     dust along the line of sight to the cluster typified by a
     \citet{SK73} law. Alternatively, a deviation from
     that law might indicate the presence of excess polarization
     with a different wavelength dependence as would be expected,
     for example, in the case of light scattered by dust particles.

\end{enumerate}

\section{The Observations \label{s_Obs}}

Linear polarization observations were carried out in September or October
during 1997, 1998, 1999, 2000 and 2001 runs with the 2.15 m telescope at
the Complejo Astron\'omico El Leoncito (San Juan, Argentina) and a
rotating plate polarimeter. This instrument, is an improvement of the
original designs known as MINIPOL and VATPOL \citep*[see][]{MA84} and takes
advantage of two high throughput photocells \citep{CASPROF}. Standard
stars for polarization were observed in all runs aiming at determining
both the instrumental polarization (less than $0.02 \%$ in all the $\ubvr I$
filters) and the zero point of the polarization angle system.

Visual polarizations were measured using a $17\arcsec$ diaphragm centered
at the position of bright red giants, i.e., the output values correspond
not only to the stars but also to the fainter stellar background
within the diaphragm. No attempt was made to correct for a local
background, since it varies in a strong way with position within the
cluster and, instead, a blank region located $45\arcmin$ to the south of
the cluster center was repeatedly observed during each night in order to
determine the temporal variation of the sky polarization. Integration
times ranged from 10 to 15 minutes per observation. The rms error of the
polarization, estimated according to \citet{MA84}, as a
function of magnitude, can be well represented with $\sigma P_V\,(\%) =
0.015 V - 0.117 $, where $V$ is the magnitude corresponding to the total
brightness within the $17\arcsec$ diaphragm. For the whole sample we
obtain a median rms error $\sigma P_V = \pm 0.045\,\% $, that leads to a
typical uncertainty of $\sim \pm 3^\circ$ in the polarization angles.
Corrections to the observed polarizations due to noise bias were
estimated following \citet{CLST} and neglected since, in the worst cases,
 resulted smaller than $0.01 \%$.

The bright-star-free regions are approximately located along the
isophotal contour of the cluster  corresponding to a major semi-axis
of $110\arcsec$, i.e., some 4.6 core radii, adopting a core radius of
$24\arcsec$ \citep*[see, for example,][]{HGG00}. These regions were
also observed with a $17\arcsec$ diaphragm and integration times
of 15 min each, yielding a typical polarization error of $\sigma P_V= \pm
0.05 $.

Polarimetry of the cluster nucleus was obtained through $UBV$ (Johnson)
 and RI (Kron-Cousins) filters during the 1997 run. In this case, the
observations were carried out through diaphragms with diameters of
$17\arcsec$, $33\arcsec$, and $45\arcsec$ and integration times from 5 to
15 minutes for each observation.  In total, we obtained eight integrations
in the $V$ and $R$ bands and five for each of the remaining
filters. Neutral filters were added for the $VRI$ observations in order to
avoid saturation due to the brightness of the cluster nucleus.
                         
Sky polarization measurements for the bright-star-free regions and for
the nucleus were also secured at the blank region $45\arcmin$ to the
south of the cluster.

Approximate $V$ magnitudes were determined for the total brightness
observed through the $17\arcsec$ diaphragm by means of $V = v_i + C_V $,
where $v_i$ is the instrumental visual magnitude (corrected by
atmospheric extinction) and $C_V= 0.19 \pm 0.05$ a constant derived by
means of 19 stars included in the photometry by \citet{LEE77}. These stars
are far from the cluster center and not affected by crowding. No color
term was added on the basis that most of the program stars have color
indices within the color range covered by the calibrating stars.

\section{The Observed Polarizations \label{s_OP}}

\subsection{The Non Variable Stars}

%\begin{subsection}

{Figure 1} shows the finding chart for 53 stars (not reported
as variables) located within $4\arcmin$ from the cluster center, and
 whose observed $V$ polarizations are presented in Table~\ref{Table 1}
 and shown on the sky in Figure~\ref{Figure 2}(a). That table gives the star
identification number, coordinates (J2000), the approximate $V$ magnitude
measured within the $17\arcsec$ diaphragm, the observed linear visual
polarization, associated error, and polarization angle (N to E).  These
data are followed by the star identification and membership probability
(when available) from the proper motion study by \citet{TU92}, as well as
the identification, $V$ magnitudes and $(\bv)$ colors, from \citet{CF78}
and, in the last column, the heliocentric radial velocities from
\citet{MAYOR}.
%---------------------- Fig. 1 ---------------------------------
\begin{figure}[!tbh]
\includegraphics[width=0.48\textwidth]{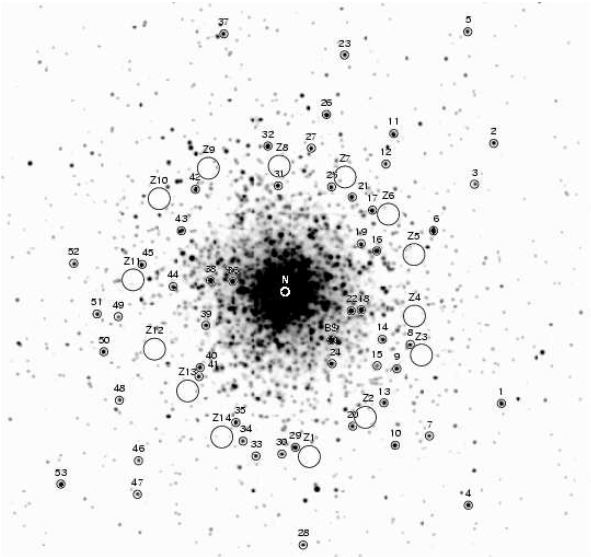}
\caption{Identification chart for the observed giant stars in
47\,Tuc. North is up, East to the left. The bright early post-asymptotic
star is labeled as BS (see text). Larger circles correspond to the
bright-star-free regions.
\label{Figure 1}}
\end{figure}
%---------------------------------------------------------------

%---------------------- Fig. 2 ---------------------------------
\begin{figure}[!tbh]
\includegraphics[width=0.48\textwidth]{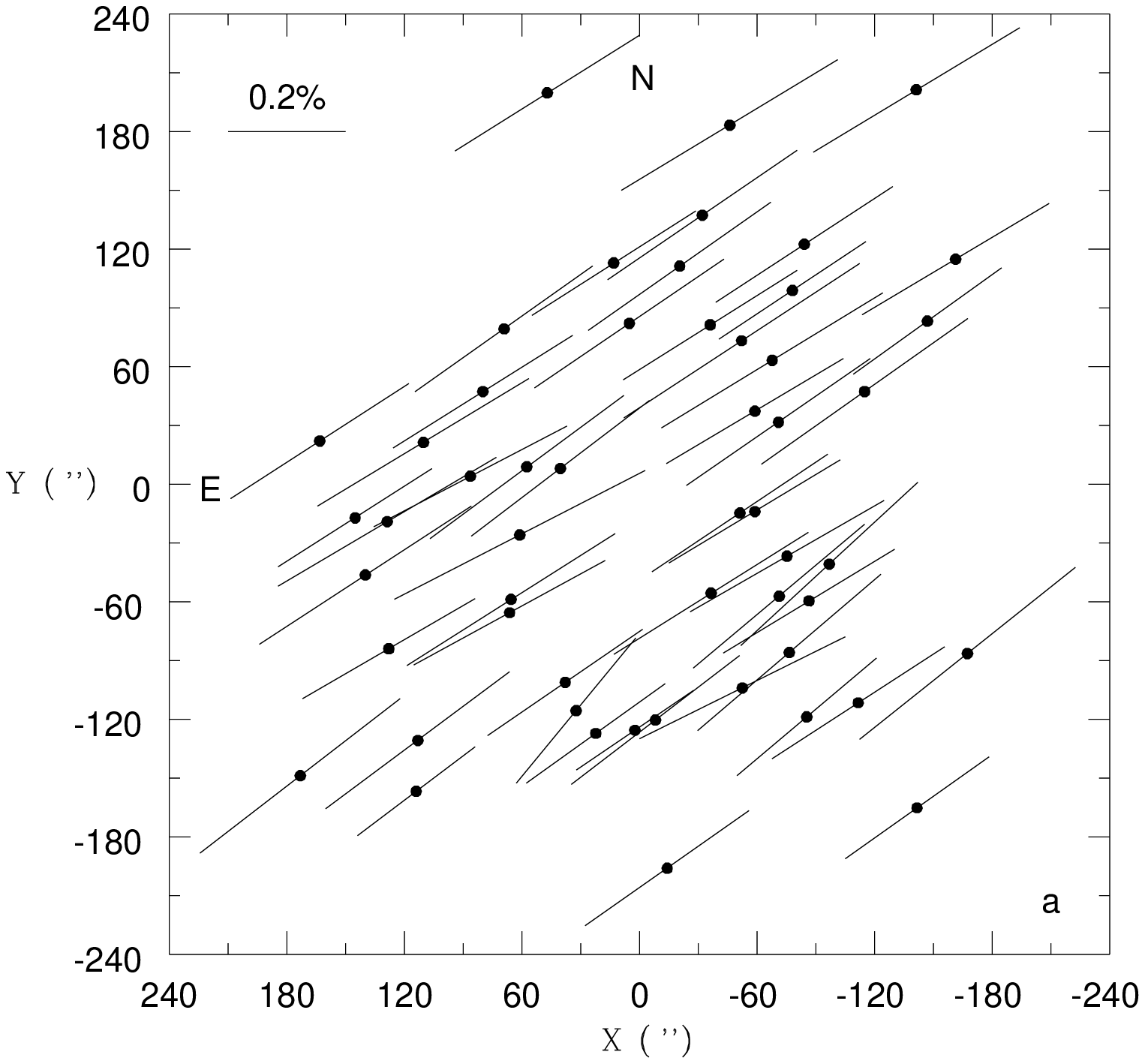}
\includegraphics[width=0.48\textwidth]{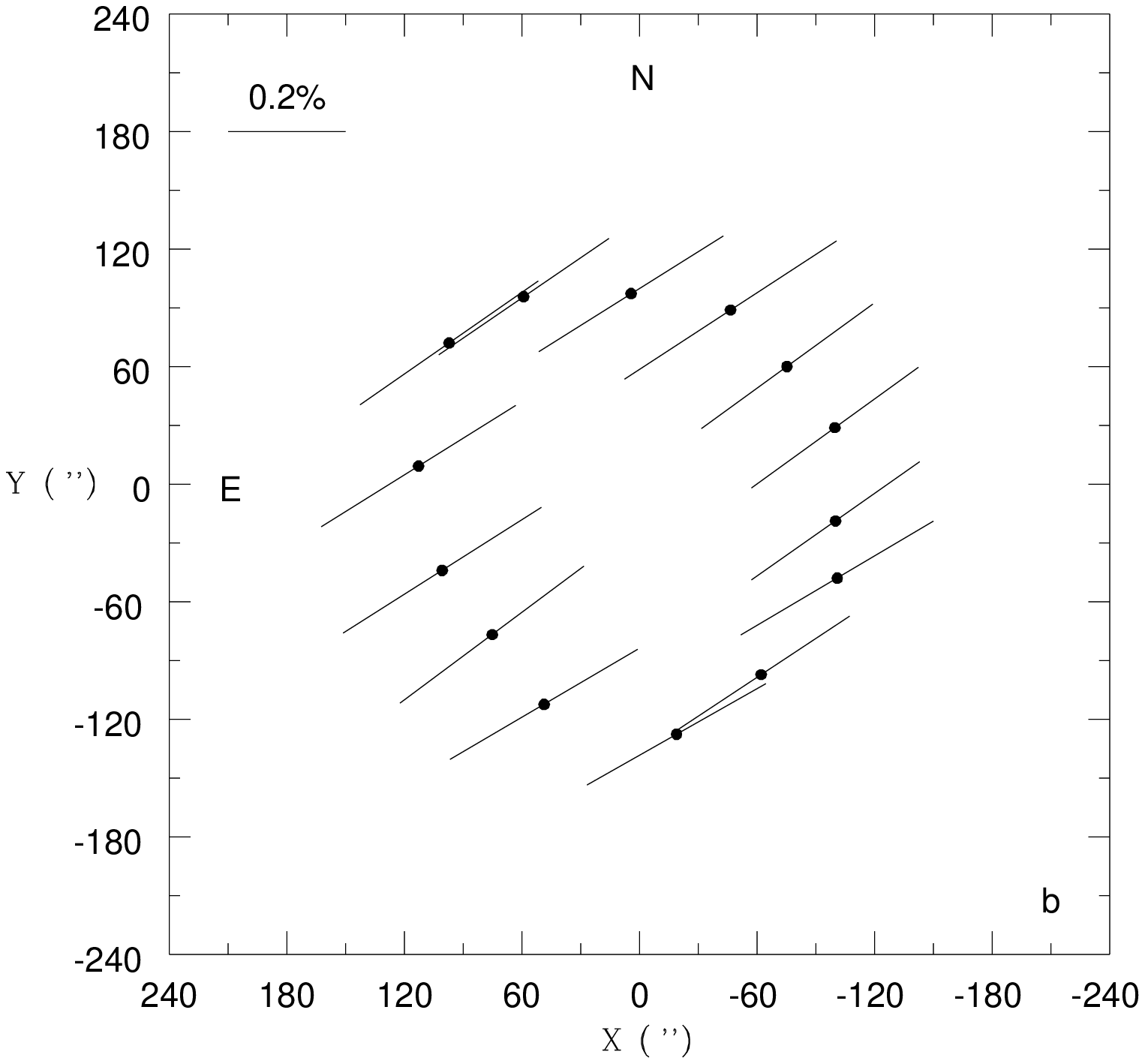}
\caption{Visual polarization vectors for the 47\,Tuc (non-variable) red
giants projected on the sky (2-a) and for the bright-star-free regions
(2-b).
\label{Figure 2}}
\end{figure}
%---------------------------------------------------------------

From the 53 stars listed in Table~\ref{Table 1}, 44 seem to be bright red
giants and cluster members, according to their position in the $V$ vs.\
$(\bv)$ diagram. This number increases to 46 by adding one star
with high membership probability from its proper motion (star 43 with no
photometry available), and by including an early type post-asymptotic star
labeled as BS by \citet{LLOYD} and UIT7 by \citet{OC97}; also see
\citet*{DI95}. For this particular star we obtain $P_{V}=0.37~\% \pm 0.02$
and $\theta_V=123.1^\circ$.  In turn, 39 of the observed stars have radial
velocities that are consistent with membership and denote the cluster
rotation (when plotted against their linear coordinates along the major
axis of the cluster; see \citet{MAYOR2}).

We note that, even though six stars have low membership probability from
their proper motions ($mp$ index lower than 0.2), neither their position in the
c-m diagram nor their polarization values exhibit significant differences
with other stars considered as cluster members and, for this reason,
were kept within the sample.

Table~\ref{Table 2} extends the polarization observations to other 14
stars located at angular distances between $4\arcmin$ and $20\arcmin$ from
the cluster center. Column 2 and 3 in this table give only approximate
rectangular positions with respect to the cluster center (X and Y
positive towards the East and North, respectively).
 All these stars have two color photometry given by
\citet{LEE77} indicating that they are probable cluster members. One
of them, L1427, as discussed later, seems to be associated with a
 long tail-like structure detected on the IRAS raster image presented
 by \citet{G88}.

\subsection{The Variable Stars}

The observed polarizations for the variable stars are given in
Table~\ref{Table 3}. The identification numbers come from
\citet{SAWYER}. One of these objects (V4) shows discrepant values when
comparing the 1997 and 1999 data; both sets are given in
Table~\ref{Table 3} as we cannot reject a possible variation of the
polarization. 

\subsection{The Bright-Star-Free Regions}

The polarizations corresponding to the bright-star-free regions are
 listed in Table~\ref{Table 4} and displayed on the sky
 in Figure~\ref{Figure 2}(b). That table also includes the coordinates (J2000),
approximate integrated $V$ magnitudes, and position angles on the sky
measured at the cluster center (N to E).

\subsection{The Cluster Nucleus}

The $\ubvr I$ observations of the cluster nucleus are listed in
Table~\ref{Table 5}.  The adopted coordinates for the cluster center were
(J2000) $\alpha= 00^\mathrm{h}~24^\mathrm{m}~5\fs95$ and
$\delta=-72^\circ~04\arcmin~53\arcsec$, within $2\arcsec$ from the center
coordinates estimated by \citet{GYSB}.  Even though the polarizations were
obtained using three different diaphragms, the $BVRI$ values do not show a
significant trend with aperture and then we only give the average values
for each of these bands.  The ultraviolet measures do show significant
differences in polarization with aperture (see Section~\ref{s_MPN}) and
then the individual values for each diaphragm are given.

\section{The Foreground Polarization}

The detection of intrinsic polarization requires the removal of
the foreground component arising in the interstellar dust along the line of
sight to the cluster. Estimates of the color excess $E_{(B-V)}$ produced by
this dust range from 0.03 to 0.04 \citep[see][]{HESSER}.

\begin{figure}[!tbh]
\includegraphics[width=0.48\textwidth]{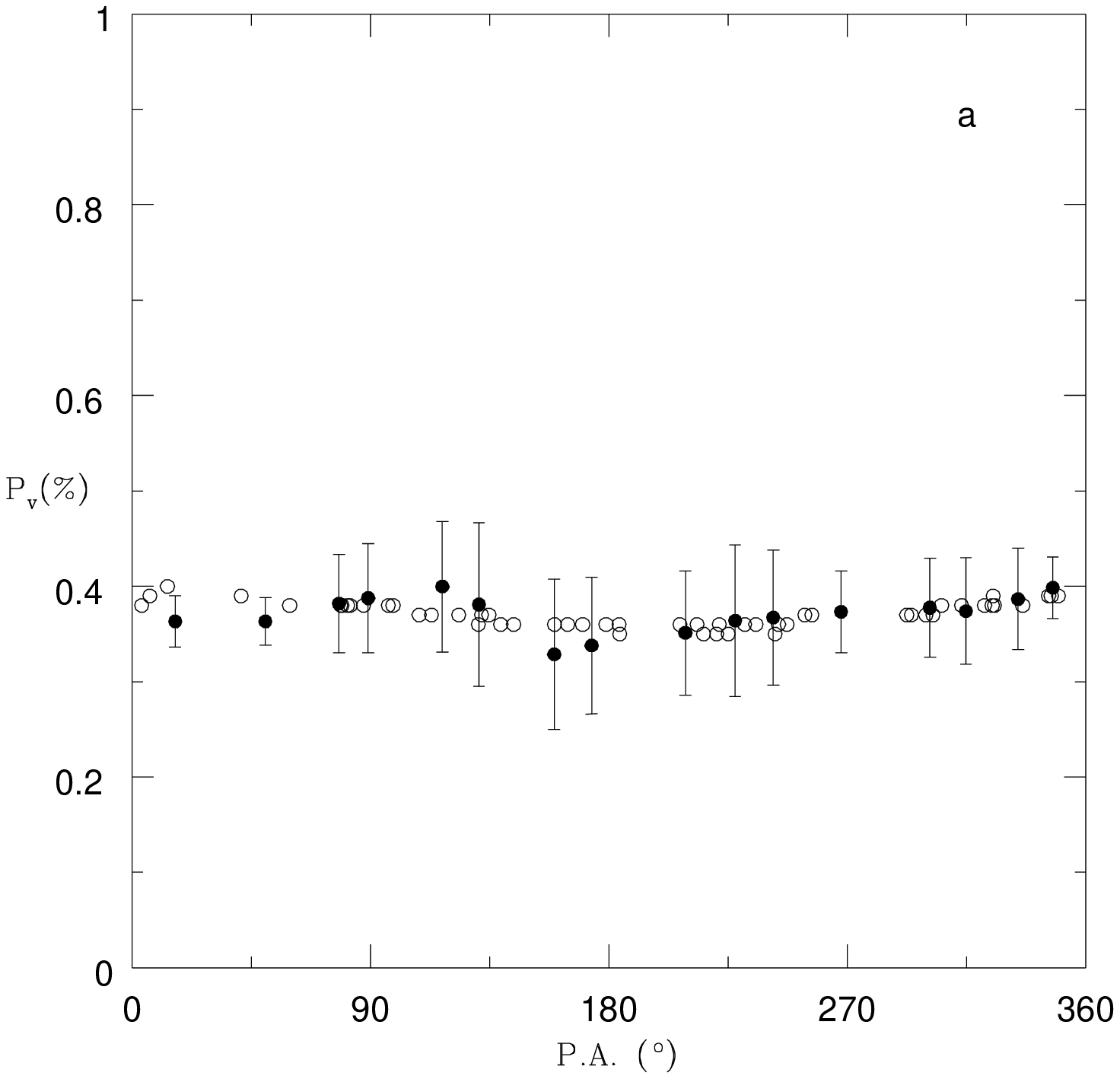}
\includegraphics[width=0.48\textwidth]{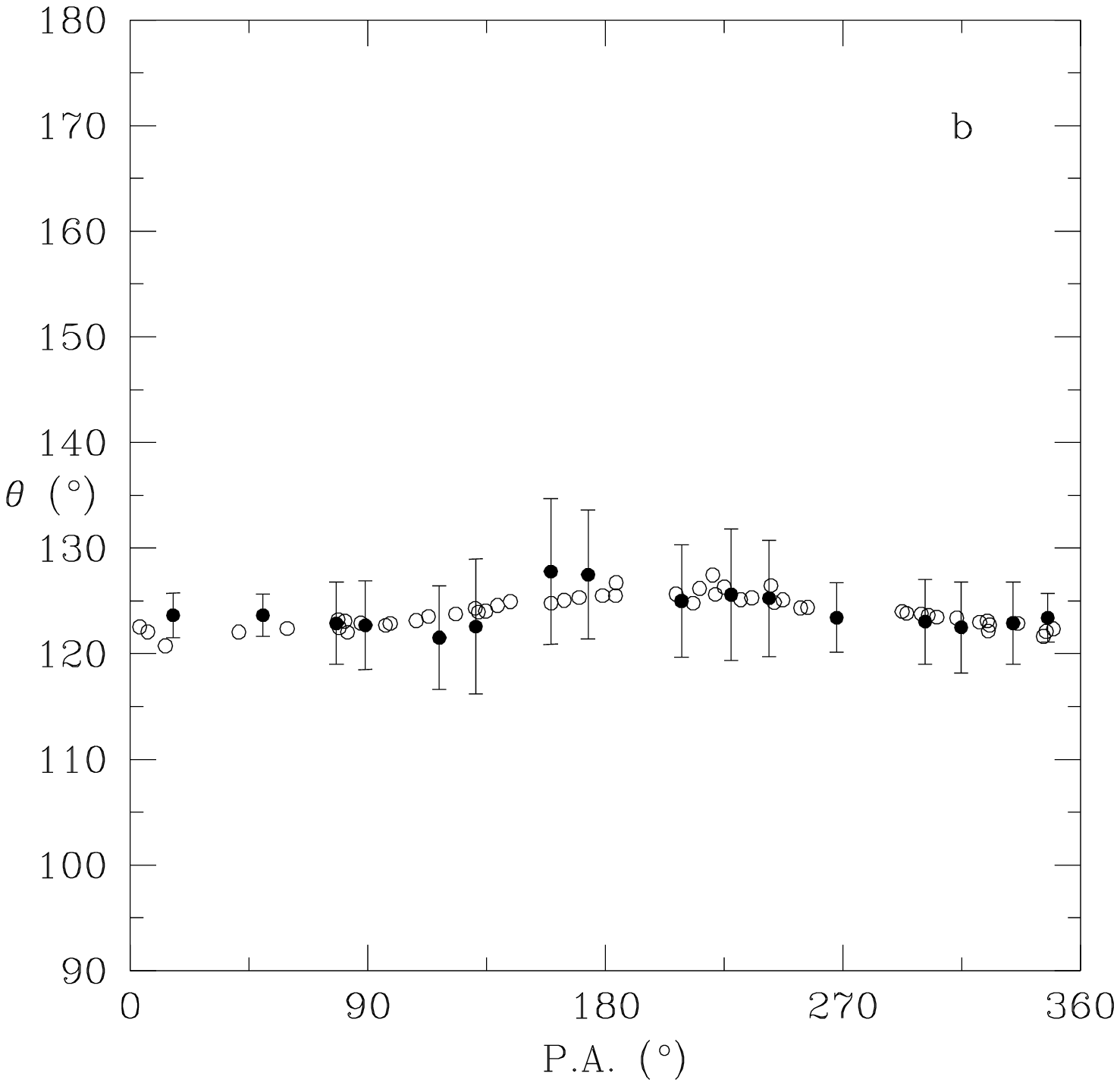}
\caption{Smoothed $P_V$ polarization (3-a) and polarization angle
$\theta_V$ (3-b) for the red giants compared with the expected polarizations
 obtained from the $U$ and $Q$ plane fittings (open circles)
as a function of position angle measured at the cluster center. The bars
represent the dispersion of the mean polarization values.
\label{Figure 3}}
\end{figure}

In order to map the foreground polarization on a scale of few arcmin,
we first used the observations of the non-variable stars listed in
Table~\ref{Table 1}. The normalized Stokes parameters for these stars were
approximated by a least squares fit to two planes defined as:
\begin{eqnarray*}
 U & = & a_\mathrm{u}\,X + b_\mathrm{u}\,Y + C_\mathrm{u}  \\
 Q & = & a_\mathrm{q}\,X + b_\mathrm{q}\,Y + C_\mathrm{q}
\end{eqnarray*}
where $X$, $Y$ are the rectangular coordinates of the stars (in arcsecs)
in a system with the positive X axis pointing towards the East, the
positive Y axis towards the North and the origin at the cluster
center. The units of the $a$ and $b$ coefficients are, then, $\%~arcsec^{-1}$
and $\%$ for the $C$ ones.
 The resulting coefficients and associated errors are:
$a_\mathrm{u}=-7.7(\pm 2.5)\times 10^{-6}$; $b_\mathrm{u}=-4.95~(\pm 2.0)
\times 10^{-6}$; $C_\mathrm{u}=-0.34 \pm 0.02$ and $a_\mathrm{q}=-1.22
(\pm 1.5) \times 10^{-4}$; $b_\mathrm{q}=-2.15~(\pm 0.5) \times 10^{-4}$;
$C_\mathrm{q}=-0.14 (\pm 0.02)$,
and show a mild trend of the $U$ parameter with both $X$ and $Y$ while the
$Q$ parameter exhibits a detectable positional dependence mainly with the
Y coordinate. A comparison between the polarization predicted by the plane
fitting (at the position of each star) and a smooth average of the
observed polarizations as a function of position angle (measured at the
center of the cluster) is depicted in Figure~\ref{Figure 3}. The smooth
average polarizations were obtained by computing the mean $U$ and $Q$
values within sectors $45^\circ$ wide and adopting a step of $22.5^\circ$.
That figure shows that the plane fit is a good overall representation of
the polarization variation within $4\arcmin$ from the cluster center both
in amplitude and angle. In turn, the $C_u$ and $C_q$ values lead to
$P_V=0.37~\% \pm 0.02$~ and~ $\theta_V=123.8^\circ$ that is representative
of the interstellar polarization at the cluster center, in very good
agreement with the (unfiltered) polarimetry by \citet{MF}, who obtain
$P=0.36~\% \pm 0.09$ and $\theta=123.0^\circ$.

%-------------------------- Fig. 3 -----------------------------
%---------------------------------------------------------------

If a color excess of $E_{(B-V)}=0.04$ is adopted for the interstellar
reddening towards 47\,Tuc, and taking into account the relation between
polarization efficiency and color excess derived by \citet*{SMF75},
$P_V=9.0\, E_{(B-V)}$, the estimated polarization for the cluster center
is compatible with a situation of maximum grain alignment and then
maximum polarization efficiency (i.e., degree of polarization per color
excess or extinction in magnitudes).
 We note that the IRAS calibrated
$E_{(B-V)}$ map by \citet*{SCHLE} predicts a somewhat smaller color excess,
$E_{(B-V)}=0.03$, at the position of the 47\,Tuc center. This map shows a
very uniform color excess on an angular scale of $4\arcmin$ from the cluster center,
i.e., the polarization variation seems more associated with a change in
polarization efficiency than with a varying foreground extinction.

%-------------------------- Fig. 4 -----------------------------
\begin{figure}[!tbh]
\includegraphics[width=0.48\textwidth]{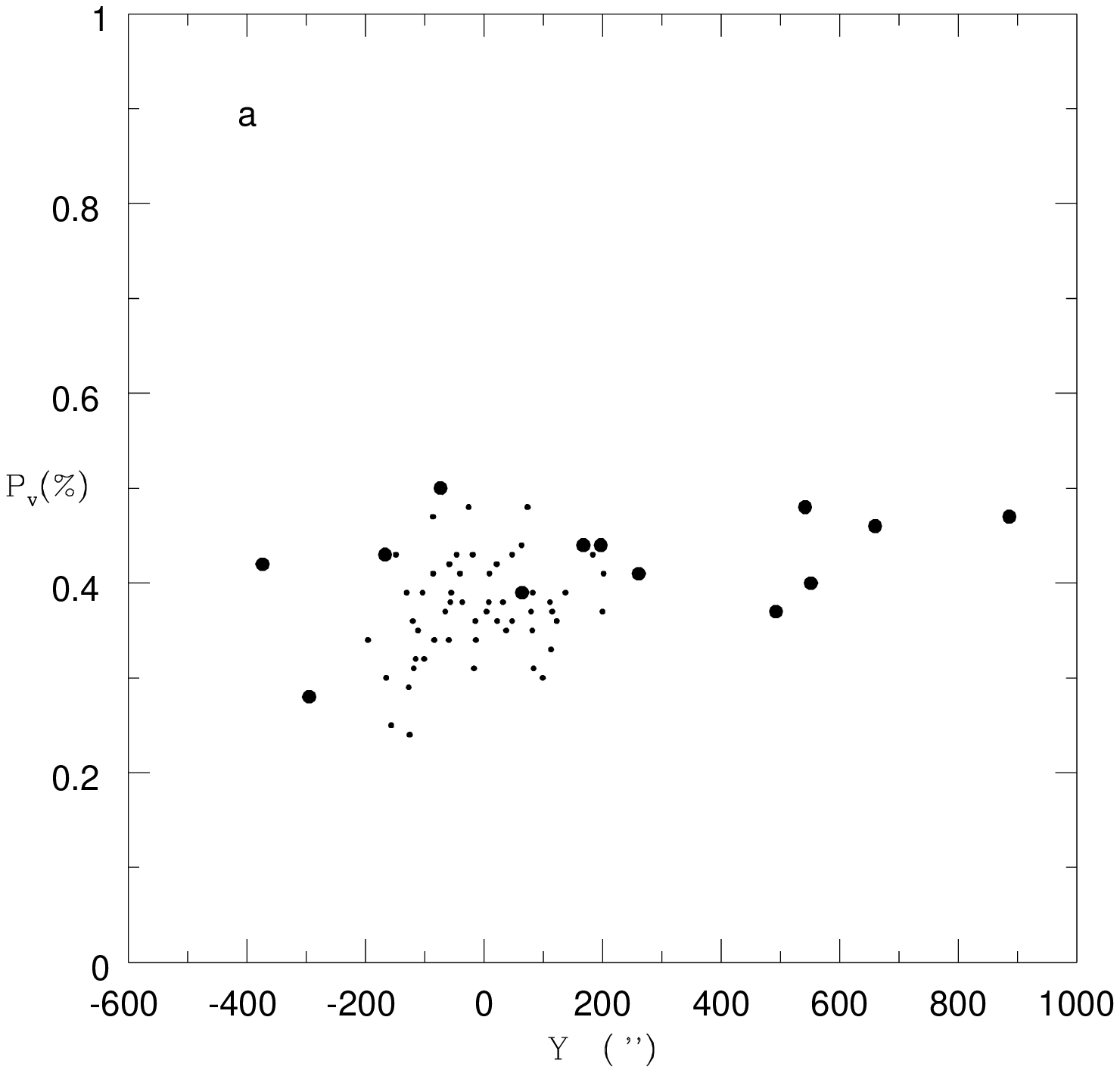}
\includegraphics[width=0.48\textwidth]{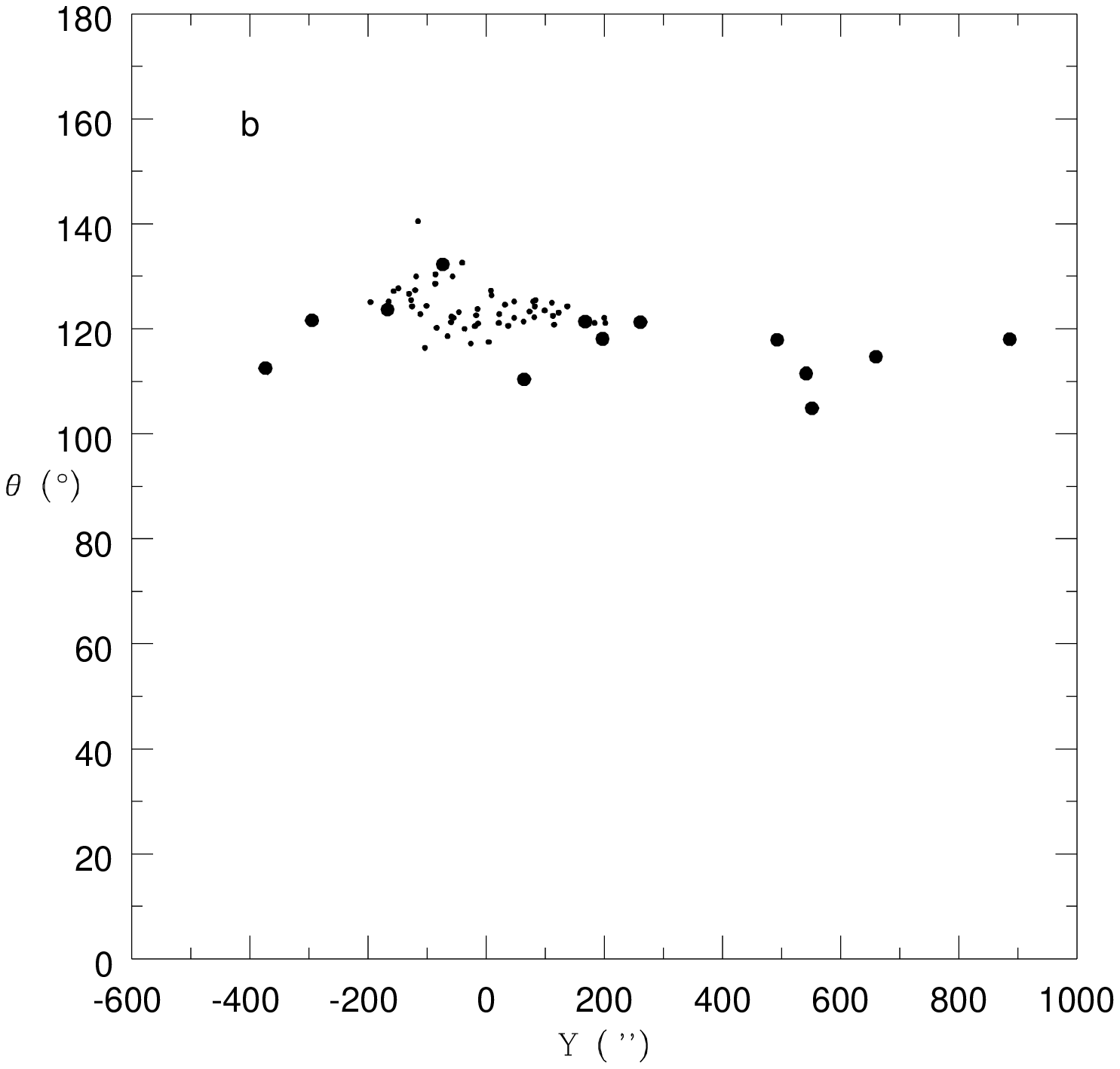}
\caption{Polarization $P_V$ vs.\ $Y$ coordinate (S-N) (4-a) and
 polarization angle $\theta$ vs.\ $Y$ (4-b) for the outer 47\,Tuc field
 (filled circles) and for the inner cluster field (dots).
\label{Figure 4}}
\end{figure}
%---------------------------------------------------------------

The polarization behavior within $4\arcmin$ from the cluster center can be
compared with that of the stars listed in Table~\ref{Table 2} (located
between $4\arcmin$ and $20\arcmin$ from the center). In this case, the
relatively small number of member stars, and the somewhat uneven
distribution on the sky, prevent a treatment similar to
that of the inner field. However, as shown in Figure~\ref{Figure 4}, the
behavior of $P_V$ for stars in this sample, and that of the inner field
stars, are compatible with an overall variation of the polarization along
the N--S direction. Thus, the observed variation within $4\arcmin$ of the
cluster center does not seem a feature just connected with the cluster
but, rather, a consequence of a variation of the polarization on a larger
angular scale on the sky.

%-------------------------- Fig. 5 -----------------------------
\begin{figure}[!tbh]
\includegraphics[width=0.48\textwidth]{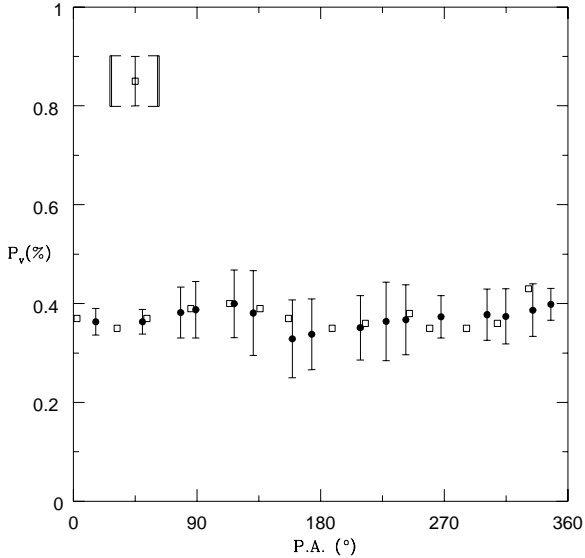}
\caption{A comparison of the smoothed star polarizations with those
corresponding to the bright-star-free regions (open squares) as a function
of position angle. Polarization samples were taken within $45^\circ$
sectors moved with $22.5^\circ$ steps. The bars represent the dispersion of the
mean polarization values. The bar in the upper left is representative
of the errors for the bright-star-free-regions.
\label{Figure 5}}
\end{figure}
%---------------------------------------------------------------

Figure~\ref{Figure 4} does not include the star L1421 (listed in
Table~\ref{Table 2}) since, as mentioned in Section~\ref{s_Obs}, this
object appears associated with an elongated infrared structure and, after
removing the interstellar component, shows an intrinsic polarization
$P_V=0.38~\%$ and $\theta_V=80.0^\circ$.

Figure~\ref{Figure 5}, in turn, shows the variation of the polarization
observed in the bright-star-free regions, as a function of their
position angle, compared with the smooth average polarization of
the non-variable stars. The very good agreement, in this case, also lends
support to the fact that most of the observed variation of the polarization
is due to the foreground interstellar component.

\section{The Residual Polarizations}

%-------------------------- Fig. 6 -----------------------------
\begin{figure}[!tbh]
\includegraphics[width=0.48\textwidth]{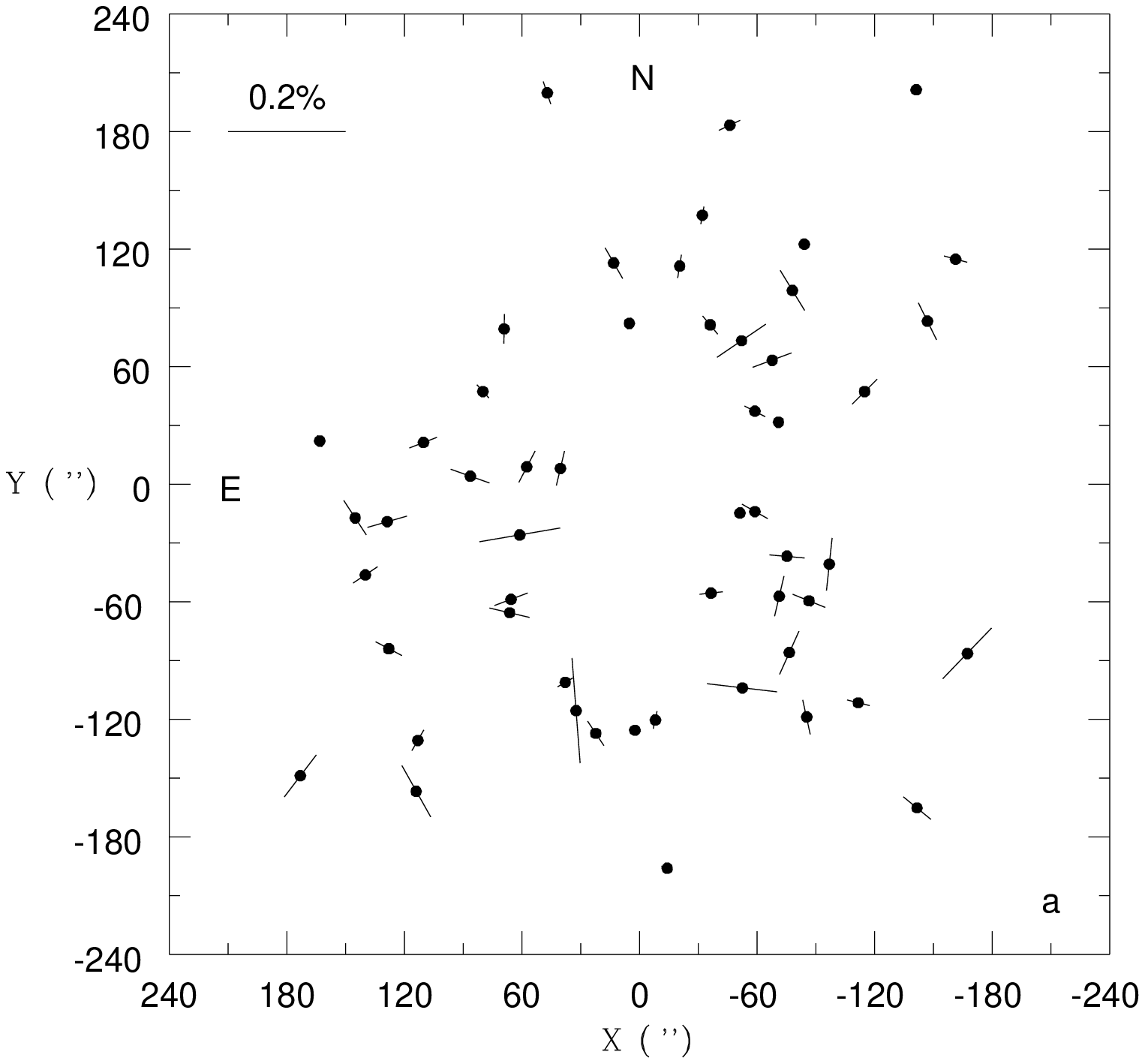}
\includegraphics[width=0.48\textwidth]{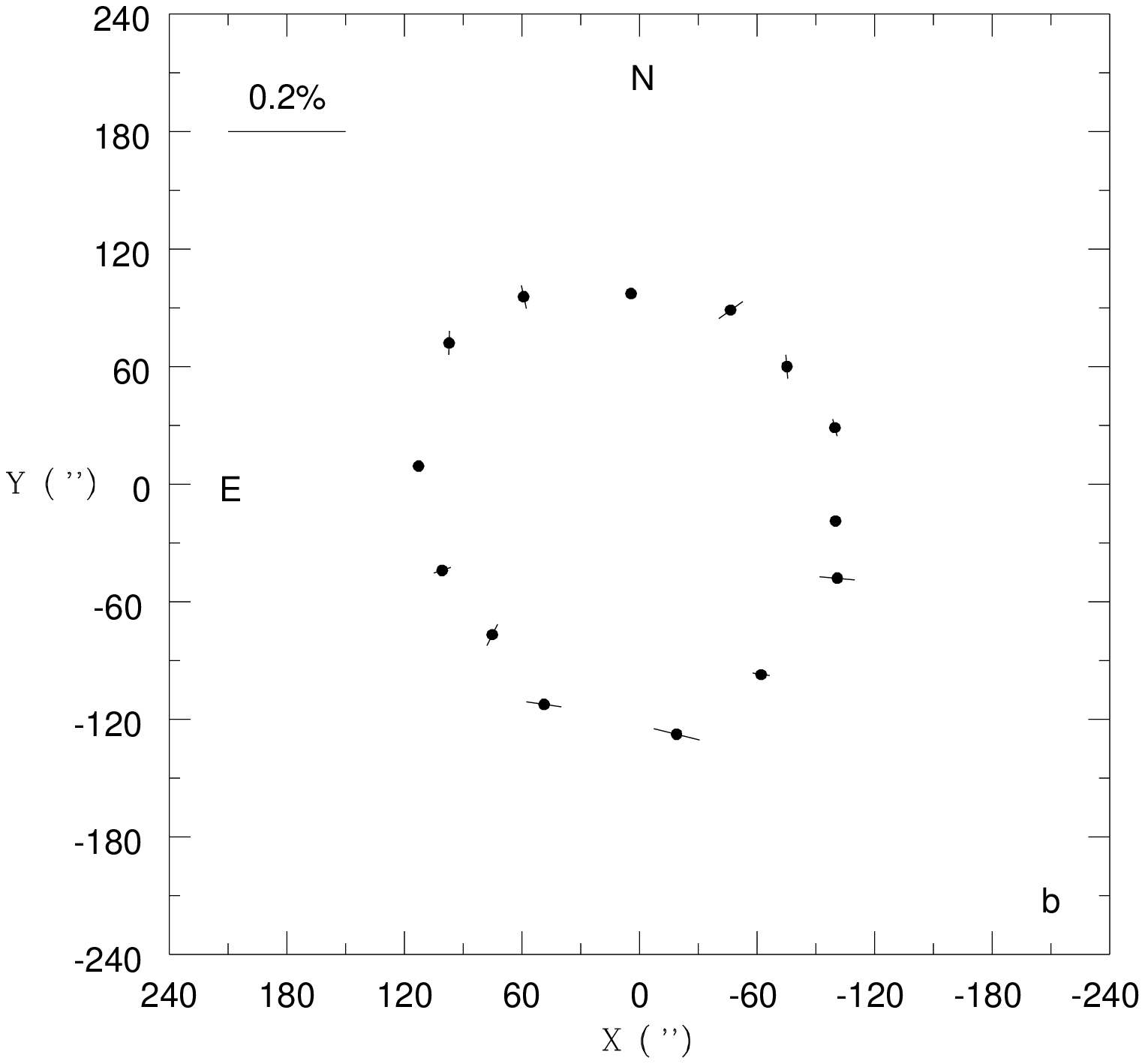}
\caption{Residual polarizations (i.e., after removing the foreground
interstellar component) for the red giants (6-a), and for the
star-free-regions projected on the sky (6-b).
\label{Figure 6}}
\end{figure}
%---------------------------------------------------------------

%-------------------------- Fig. 7 -----------------------------
\begin{figure}[!tbh]
\includegraphics[width=0.48\textwidth]{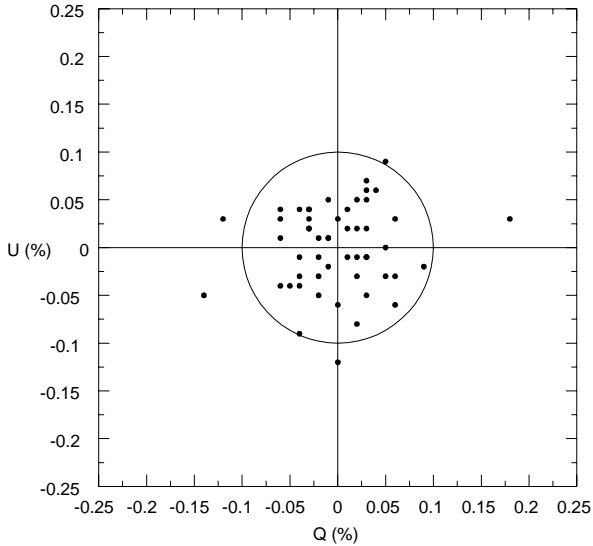}
\caption{$U$ vs.\ $Q$ Stokes plane for the non variable red giants
 after removing the interstellar polarization. The circle has a radius set
 by $2.5~ \sigma P_V$.
\label{Figure 7}}
\end{figure}
%---------------------------------------------------------------

The residual polarizations, after subtracting the foreground component
defined by the parameters discussed in the previous section, are shown on
the sky plane both for the inner field stars and for the star-free regions
in Figure~\ref{Figure 6} and, on the Stokes plane, in Figure~\ref{Figure
7}. The main features in these diagrams are: a) Most of the residual
polarizations seem consistent with observational errors, and practically,
no star exhibits values exceeding the $3\, \sigma{P_V}$ level; b) There
is no systematic angular pattern, neither for the star nor for the star-free-region
polarizations.

From these results we conclude that intrinsic polarizations
associated with dusty shells, are not detectable in most of the non-variable stars.

%-------------------------- Fig. 8 -----------------------------
\begin{figure}[!tbh]
\includegraphics[width=0.48\textwidth]{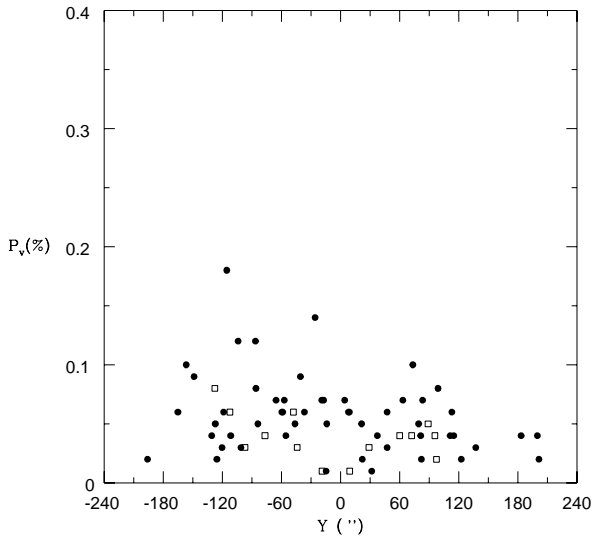}
\caption{Residual polarizations for the red giants (filled circles) and
for the star-free-regions (open squares) as a function of the $Y$
coordinate (S--N). An increasing trend is seen towards the south of the
cluster.
\label{Figure 8}}
\end{figure}
%---------------------------------------------------------------

Even though these residual polarizations are not large compared with the
errors of the observations, a plot of these values as a function of the Y
coordinate (along the N--S direction), displayed in Figure~\ref{Figure 8},
shows a systematic trend in the sense that the larger polarizations are
located towards the south of the cluster center. Eight (out of nine stars)
with residual polarizations at a $2\, \sigma P_{V}$ level or larger,
appear in that position. This trend is also seen in the two southernmost
star-free regions (also plotted in Figure~\ref{Figure 8}). Although this
effect is marginal, and deserves higher precision polarimetry in order to
be confirmed, it could be an indication of the presence of small
amounts of scattered light. As the polarimeter aperture includes an
important fraction of background light the nature of the observed
polarization, i.e., if the residual polarizations are originated by the stars or by
the background light, is not clear. However, as most of the
bright-star-free regions do not show detectable polarization excesses,
we are inclined to believe that they could arise
in low optical depth dust locally illuminated (but not
necessarily originated by) some stars in the southern region.
In this case, and assuming that the background light within the diaphragm
 is only polarized by the foreground dust along the line of sight to the
 cluster, the residual polarizations should be increased by a factor
 $10^{0.4\delta m}$, where $\delta m$ is the difference between the star
 and the total V magnitudes listed in Table 1. In the most extreme
 cases such a correction leads to polarization excesses as large as
 $0.35~\%$.

The fact that the proper motion of the cluster in its own LSR, as shown in
Table 3 in \citet{KG95}, points towards the north, may give some
basis to speculate about the nature of the observed spatial asymmetry of
the residual polarizations seen in the opposite direction(see the
 Discussion section).

\section{The Variable Stars}

%-------------------------- Fig. 9 -----------------------------
\begin{figure}[!tbh]
\includegraphics[width=0.48\textwidth]{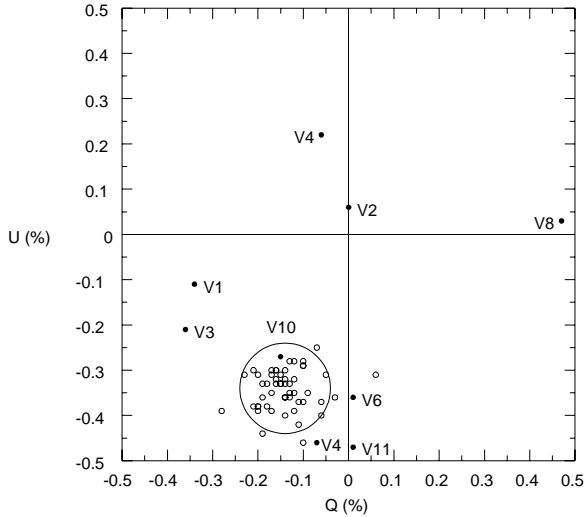}
\caption{Stokes $U$ vs.\ $Q$ plane for the variable stars (filled circles).
The observations corresponding to V4 are indicated in a separate way.
Non variable red giants are shown as open circles. The circle has
a $2.5\sigma P_{V}$ radius and is centered at the $U$, $Q$ values corresponding
to the mean interstellar polarization along the line of sight towards
the cluster. V13 falls out of the limits of the figure.
\label{Figure 9}}
\end{figure}
%---------------------------------------------------------------

The observed $U$ and $Q$ parameters for the group of nine red luminous
variables in our sample are displayed in Figure~\ref{Figure 9}. This
figure also includes a circle with a $2.5\, \sigma P_{V} $ radius that
contains, practically, all the non variable stars (also plotted). The
 circle is centered
at the U and Q values that are representative of the interstellar
 polarization ($C_u$~ and~$C_q$ parameters in Section 4). Except for the variable
star V10, all the remaining ones fall outside the error circle and then
exhibit a significant excess or intrinsic polarization.  For seven of
these stars, the excess polarizations, not corrected by the
background dilution effect mentioned in the preceding section, range from $0.15\,\%$ to
$0.70\,\% $ and show no preferential alignments (but see the case of the
comparatively highly polarized V13 below).

The fact that most of the variable stars show intrinsic polarizations
agrees with a previous tentative conclusion by \citet{OR97} who point out
that ``...in the circumstellar environment of V1, V2, V3 and V4 we
have some indication of intrinsic polarization...''. In the particular
case of V1, these authors also report a detection at $10\, \micron$
attributed to dust emission. The presence of dust around V3, in turn, is
supported by the high $12\, \micron$ to $2\, \micron$ flux ratio found by
 \citet{RAJO01}. These last authors also mention a marginal detection
of dust emission around V11 based on the same flux ratio.

The variable star V13 is an interesting object (although its membership
into the cluster remains to be confirmed). Its intrinsic polarization,
 after removing the foreground interstellar component,
reaches $ P_{V}=2.0~(\pm 0.15) \%$ and a polarization angle
$\theta=123.9^\circ (\pm 3^\circ) $, i.e., remarkably similar to that of the
foreground polarization itself.  This coincidence requires a further
analysis in order to clarify if it happens just by chance or,
alternatively, if the alignment mechanism of the foreground dust and the
one that controls the circumstellar polarization in V13 are coupled in
some way. Since this last polarization is thought to arise in scattering
(as supported by preliminary $B$ band observations that show an
increase of the polarization towards shorter wavelengths in V13) while the
foreground dust polarization is associated with large scale magnetic
fields, the physics behind that hypothetical coupling is not clear. If dust
alignment by these magnetic fields is responsible for the observed
polarization then, in a case of maximum polarization efficiency, a color
excess $E_{(B-V)}=0.25$ would be required. This excess means an interstellar
reddening some seven times larger than that estimated for the 47\,Tuc
field.

\section{Multicolor Polarization Observations of the 47\,Tucanae Nucleus
\label{s_MPN}}

\begin{figure}[!tbh]
\includegraphics[width=0.48\textwidth]{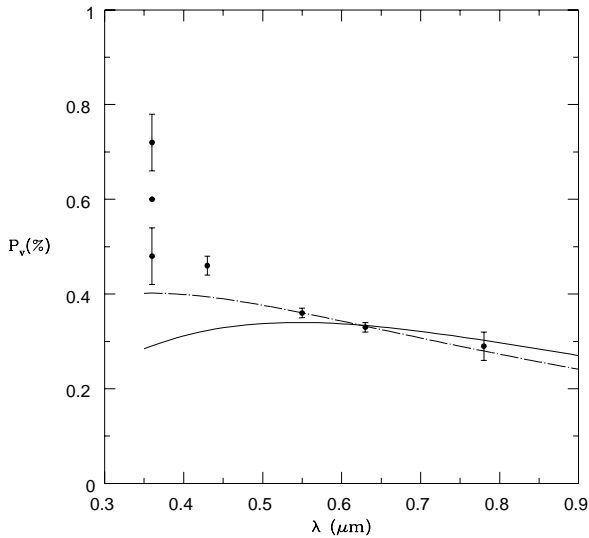}
\caption{Linear polarization observed in the 47\,Tuc nucleus as a
function of wavelength of the $\ubvr I$ bands. Three different values are
given for the $U$ filter corresponding to the $33\arcsec$ and $45\arcsec$
diaphragms and to their average. The lines correspond to two different
Serkowski law fits with $\lambda_\mathrm{max}$ of 0.55 $\micron$ (solid line)
and 0.36 $\micron$ (dot-line).
\label{Figure 10}}
\end{figure}

The $\ubvr I$ polarization observations of the cluster nucleus, listed in
 Table~\ref{Table 5}, are depicted as a function of wavelength in
 Figure~\ref{Figure 10}. Usually, the wavelength dependence of the
 interstellar polarization can be adequately fit with a \citet{SK73}
%Serkowski
curve
\begin{equation}
(P_{\lambda}/P_{\lambda_\mathrm{max}})=\exp^{-K\,
[\ln(\lambda/\lambda_\mathrm{max})]^{2}},
\end{equation}
that involves three parameters: $\lambda_\mathrm{max}$,
$P_{\lambda_\mathrm{max}}$, and $K$. Later discussions by \citet*{WILL82}
and \citet{WHIT92} have shown the existence of a linear correlation
between $K$ and $\lambda_\mathrm{max}$. In most cases, the interstellar
polarization exhibits a peak with a wavelength close to $0.55\, \micron$,
which is consistent with a selective extinction ratio $ R \sim 3.0$
\citep{SMF75} a behavior that is in contrast with the fact that the nucleus of 47\,Tuc
 shows a monotonic increase of the polarization towards shorter
wavelengths. This situation cannot be considered, in principle, as an
anomalous case as shown by some objects observed by \citet{WHIT92} (see
their figure 3) which have maximum polarization wavelengths as short as
$\sim 0.35\,\micron$. The assumption that the interstellar polarization
curve for the 47\,Tuc nucleus peaks in the $U$ band, would imply $K=0.61$
according to the K vs $\lambda_\mathrm{max}$ relation given by
\citet{WHIT92}.  However, such a curve, normalized to $P_V=0.36\,\%$,
gives a poor representation to the observed data, as shown by Figure
\ref{Figure 10} where another fit, just considering the $VRI$ values, is
also shown. Moreover, the $U$ band observations through the $
33\arcsec $ and $ 45\arcsec $ diaphragms exhibit a significant variation
in polarization amplitude and a different angle compared with those
observed for the other filters. These two features suggest that the
observed polarization cannot be entirely explained as a result of an
interstellar component typified by somewhat extreme values of $K$ and
$\lambda_\mathrm{max}$.

%-------------------------- Fig. 10 ----------------------------
%---------------------------------------------------------------

We note that the observed visual polarization for the nucleus agrees, within
the errors, with the estimated foreground polarization discussed in
Section~\ref{s_OP} then suggesting that the excess polarization in the $V$
band, if any, is very small.

\section{A Tentative Explanation for the Excess Polarization in
the 47\,Tuc Nucleus \label{s_EPN}}

The bright (and blue) post-asymptotic star referred as BS or UIT7
(see subsection 2.1) is located some $48\arcsec$ from the cluster
center and appears as a suitable candidate to be considered as the origin
of the excess polarization in the nucleus for two reasons:

a) The angle of the excess polarization, after removing the ultraviolet
interstellar polarization expected from a Serkowski curve, is close to
``tangential'' (i.e., perpendicular to the line joining the star and the
cluster center, see below) as expected if some dust in the nuclear region
were asymmetrically illuminated by the star; and b) The ratio of the star
to nucleus flux, as a function of wavelength, increases in a steep way
towards the ultraviolet then decreasing the dilution effect produced by
the light of the nucleus within a given aperture.
This effect would be reflected, as observed, by an increase of the excess
polarization in $U$ and to a lesser extent at longer wavelengths, where
that flux ratio decreases.

We quantify this polarization excess first, by adopting the average of the
polarization measures through the $33\arcsec$ and $45\arcsec$ diaphragms,
$P_U= 0.60~ \%$ and $\theta=117.9^\circ$ and second, by removing the
ultraviolet interstellar polarization. For this component, a maximum
polarization wavelength of $0.36\,\micron$ (from
Figure~\ref{Figure 10}) or $0.55\,\micron$ (compatible with a more
``normal'' behavior), with $P_{V}=0.36\,\%$ and $\theta_V=123.8^\circ$, leads
to an excess polarization in the $U$ band of $0.23\,\%$ or $0.30\,\%$,
with angles of $\theta_U= 118.9^\circ$ and $\theta_U=113.7^\circ$
respectively. These polarization angles lie within less than $20^\circ$
from the angle of the direction perpendicular to the line joining the star
and the cluster nucleus ($\sim 132^\circ$), that is, the polarization
angle expected in the case that dust were effectively illuminated by the
post-asymptotic star.

A quantitative estimate of the dust optical depth required to originate
such excess polarization requires:

a) The knowledge of the spectral energy distributions both for the star
and the cluster nucleus. The spectral type of UIT7 is B8III, with an
apparent visual magnitude $V=10.73$ according to table 1 in
\citet{OC97}, implying a temperature $T=13450^\circ\, K$ and intrinsic
colors $(\ub)_0=-0.44$~ and~ $(\bv)_0=-0.14$ \citep[from][]{CRA84}.
Alternatively, \citet{DI95} obtain a considerably lower temperature,
$T=10\,500^\circ\,K$ corresponding to $(\ub)_0=-0.20$ and $(\bv)_0=-0.01$
\citep{FL96}. As a compromise we then take the average of these indices,
$(\ub)_0=-0.32$~ and~ $(\bv)_0=-0.08$ as representative for the intrinsic
colors of UIT7. In turn, adopting a color excess $E_{(B-V)}=0.04$ towards
the cluster and a selective extinction ratio $R=3.0$, lead to interstellar
extinction corrected magnitudes $U_0=10.21$, $B_0=10.53$ and $V_0=10.61$
for the star.

For the nucleus we adopted the reddening corrected colors given
by \citet*{REED}, $(\ub)_0=0.31$, $(\bv)_0=0.82$. Within an aperture of
$33\arcsec$ we obtain $V \sim 7.50$ and then $U_0=8.51$, $B_0=8.20$~ and
$V_0=7.38$.

A comparison of the magnitudes of the star and the nucleus shows a rapid
increase of the relative brightness of UIT7 towards short wavelengths,
a fact that is dramatically shown by figure 1 in \citet{OC97}, where the
far ultraviolet luminosity of the star exceeds the integrated brightness 
of the cluster within the visual half light radius ($174\arcsec$).

b) The adoption of a relative geometry involving the cluster nucleus,
the star and the observer. The star UIT7 appears at a projected
distance on the sky of $48\arcsec$ from the cluster center. As
we have no indication about the relative position of the star and
the cluster nucleus along the line of sight, we leave this quantity
as a free parameter in our calculation.

c) The adoption of parameters for the dust grains such as size,
albedo and phase function. The fact that the excess polarization
increases rapidly towards the ultraviolet, and favors a
$\lambda^{-4}$ type of dependence rather than a $\lambda^{-1}$ law that
characterizes ``normal'' interstellar grains, indicates
that the dust scatterers might be small compared to the light
wavelength. As a first approach, we then adopt highly reflective
particles (i.e., unity albedo) and a Rayleigh-like phase function.

d) The size, spatial dust distribution and optical depth of the
 cloud. The behavior of the ultraviolet polarization, that increases
from the $17\arcsec$ diaphragm towards a maximum at $33\arcsec$ and
then decreases at $45\arcsec$, suggests that most of the polarization
arises from within the second aperture and then we adopt an indicative
cloud radius of $16.5\arcsec$ (on the sky). The spatial dust density
was represented with a function consistent with a King profile with
$r_c= 24\arcsec$, truncated at the indicative radius mentioned above,
and an optical depth (through the whole cloud diameter) $\tau_U=0.5$.
This value corresponds to a weighted optical depth of $\sim 0.3$ for
the whole cloud that may be somewhat large for an optically thin
approach and would require depolarization corrections in a more
rigorous model.

\begin{figure}[!tbh]
\includegraphics[width=0.48\textwidth]{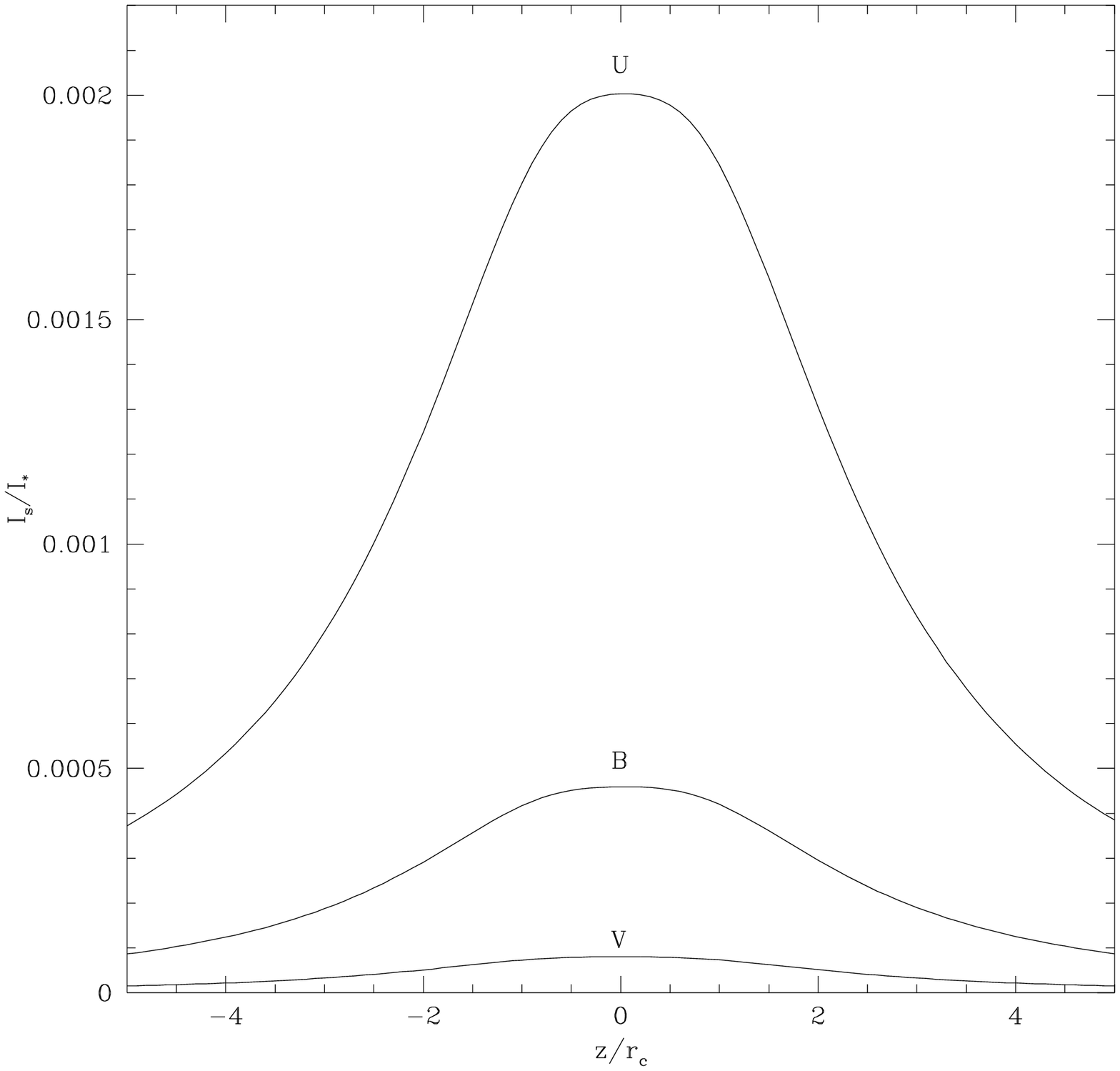}
\includegraphics[width=0.48\textwidth]{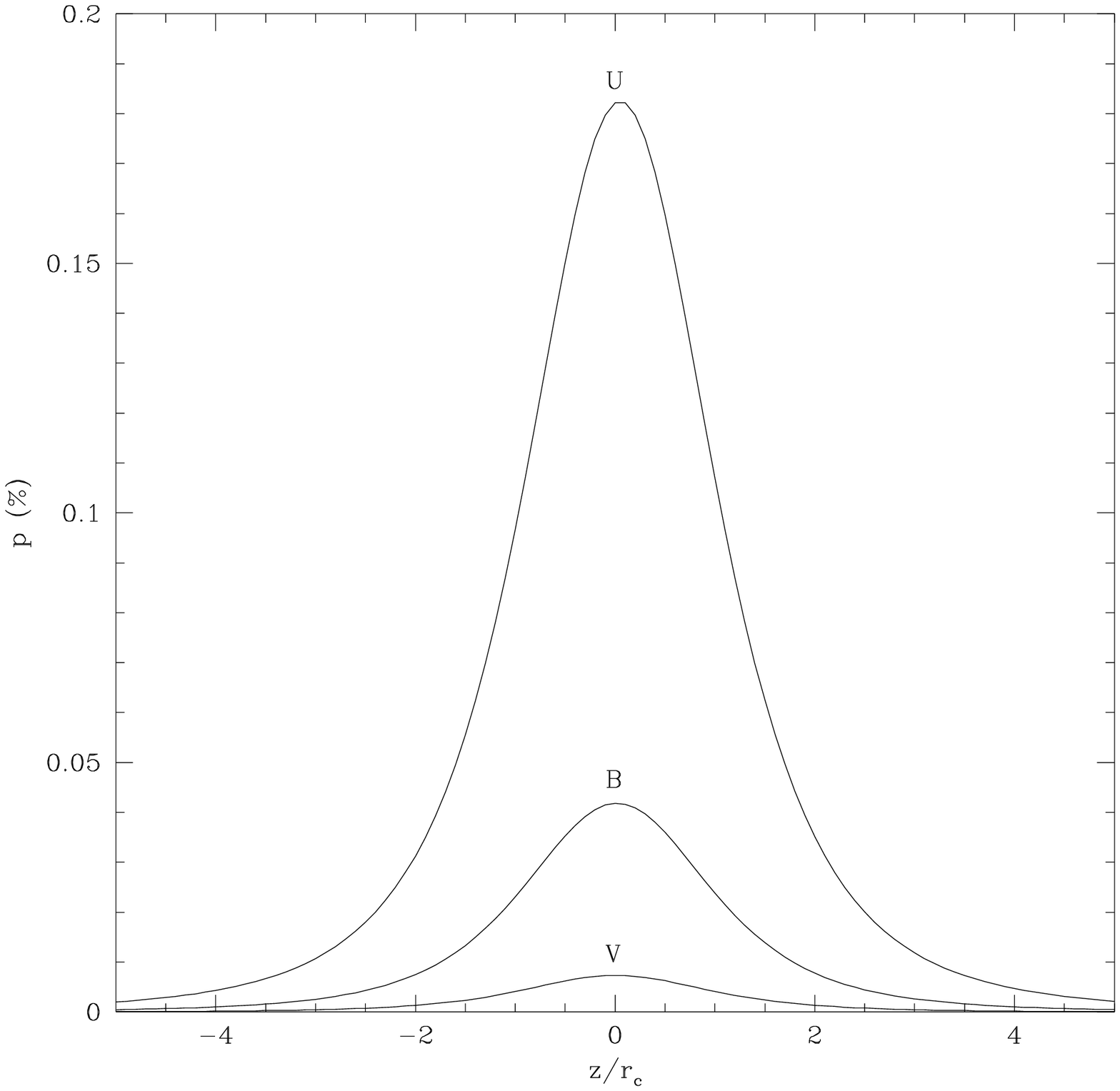}
\caption{Star (UIT7) to nucleus ($33\arcsec$ diaphragm) flux
ratio in the $UBV$ bands (11-a) and resulting polarization due to an
spherical dust cloud at the cluster nucleus with $\tau_{U}=0.5$, (11-b)
as a function of the coordinate along the line of sight (in core
radius units) and assuming Rayleigh scattering.
\label{Figure 11}}
\end{figure}

Figure~\ref{Figure 11} depicts the results, obtained with all the
mentioned assumptions, for the ratio of scattered light, arising in
the illuminating star UIT7, to that of the integrated stellar light of the
nucleus in the UBV bands, as a function of the ``depth'' coordinate, i.e.,
the position of the star along the line of sight in core radius units (11-a)
and the resulting polarization from a single scattering approach, \citep[and
equations from][]{VdH} (11-b).

%-------------------------- Fig. 11 ----------------------------
%---------------------------------------------------------------

This figure indicates that the order of the observed excess polarization
can be originated by a relatively low optical depth dust cloud centered at
the cluster nucleus and illuminated by the star UIT7. We note that, if
 this cloud is mixed with the stars
and also illuminated by the cluster radiation field, its effect on the
integrated cluster light profile will be considerably decreased. The adoption
of models like those discussed by \citet*{WITT} show that the central
profile brightness would be dimmed between 0.16 and 0.25 magnitudes
in the $U$ band (for an isotropic scattering or a Henyey-Greenstein
phase function typified by an anisotropy factor $g=0.7$) and less than
0.1 magnitudes in the $V$ band.

The estimate of the mass associated with such a cloud is dependent on the
nature of the dust grains. Following \citet{KW78} and adopting a
cloud radius of 0.36 parsecs, corresponding to an angular radius of
$16.5\arcsec$ at a distance of 4.7 kpc \citep[from][]{HESSER}:
\begin{equation}
M_\mathrm{d}=0.18\,\rho\,(\tau)\,(1/Q)\,(r_\mathrm{g})\,M_\sun
\end{equation}
where $\tau$ is the average optical depth of the spherical cloud,
 $\rho$ is the density (in g cm$^{-3}$) of a dust grain; $Q$, its extinction
efficiency ; and, $r_\mathrm{g}$ its radius in microns. For example, the
 adoption of astronomical silicate grains
\citep{DRL} with radii in the order of a few hundred \AA~ and $\tau_U=0.3$,
 leads to a total dust mass of $\sim 0.05 M_\sun$. This mass is well above
 the estimates coming from far infrared observations (see the Discussion
 Section) and includes the uncertainty connected with the fact that
 interstellar grains may not be an adequate approximation to circumstellar
 dust (if this is the origin of the dust particles producing polarization).

\section{Cloud Candidates in the 47\,Tuc Nuclear Region}

An inspection of HST archive multicolor images show the presence of
several dark patches within $30\arcsec$ from the cluster nucleus. One of
these patches, that exhibits a triangular shape ($5\arcsec \times
3\arcsec$), appears at $12\arcsec$ from the cluster center and at a
position angle of $120^ \circ$.  Two polarimetric measures were obtained
at the position of this dark patch through a $5\arcsec$ aperture diaphragm
during the 2000 and 2001 runs. These observations were made with the $V$
filter, and integration times of 300 secs each, yielding $P_V=0.42 \pm
0.02 \%$ and $\theta=123.2^\circ \pm 2$. This polarization is larger, by
some $20\,\%$, compared to that of the nucleus and suggests the existence of
excess polarization associated with the dark patch.  We note that the
geometric shape of the patch is not matched by that of the circular
diaphragm. For this reason, a relatively large amount of light
coming from its peripheral regions may produce a dilution effect on the
intrinsic polarization eventually arising in the dark patch itself.

\section{Discussion}

The analysis of the visual polarization of both the non-variable
giant stars and of the bright-star-free regions within $ 4\arcmin$
from the cluster center shows a variation that can be attributed to the
foreground dust along the line of sight to 47\,Tuc. This variation
is consistent with the polarizations observed on a larger angular scale,
i.e., stars between $4\arcmin$ and 20 $\arcmin$ from the center.
An inspection of the $E_{(B-V)}$ IRAS calibrated map \citep{SCHLE}
over the same region, however, does not exhibit a systematic change
of reddening indicating that the observed polarization trend is more
likely produced by a variation of the foreground polarization efficiency
rather than by a variable optical depth of the foreground interstellar
dust.

After a vectorial removal of the foreground polarization most of the
residual values corresponding to the non variable stars are small,
indicating that they are probably reflecting just the observing
errors. However, there remains a detectable trend in the sense that the
larger residuals are seen towards the south of the cluster, a direction
that is opposite to that of the cluster proper motion.  \citet{KG95} have
suggested that the X-ray structure observed around 47\,Tuc could be
originated by the cluster motion and the interaction of its gas with the hot galactic
halo. The same mechanism, i.e.  ram pressure on dust grains, may lead to
the formation of a tail-like structure or a preferential direction along
which dust grains leave the cluster \citep*{PELLIZZA}.  The spatial
velocity components of 47\,Tucanae given in table 3 of \citet{KG95}, based on
absolute proper motions derived by \citet{TU92} and \citet{CUD}, indicate
that such a tail would be located towards the south of the cluster and
receding from the sky plane with an angle of about $55^\circ$. The
 lack of a regular pattern of the residual polarizations
observed in some stars located towards the south of the cluster center
might be explained as a consequence of optically thin dust locally
illuminated by the stars themselves, or combined with back-scattering of the
cluster radiation as a whole. Back-scattering properties for galactic
cirrus have been discussed, for example, by \citet*{JAKOB}. This
possibility is also worth of further analysis given the existence of a
diffuse far-ultraviolet background, detected on UIT observations by
\citet{OC97}, which has a spatial scale length that is significantly different from that
of the 47\,Tuc stars. The dust scattering explanation was considered by
these last authors and, in principle, ruled out due to the large dust mass
required if ``canonical'' dust grains were present and mixed with the
stars.

Another argument that may also support the existence of
a tail-like structure comes from the appearance of the elongated structure
detected on IRAS $12~\micron$ raster images of the central regions of 47\,Tuc
by \citet{G88} (see their Figure 1-b). This detection would imply relatively 
large dust equilibrium temperatures for an extended source. However,
other radiation mechanisms, as for example, very small particles transiently
heated by absorption of single photons \citep*{SELLGREN}, may provide
an explanation consistent with the polarimetric results discussed
in Section~\ref{s_EPN} that also favor the presence of small particles.

There are other elongated or
cirrus-like structures in the IRAS image. As mentioned before, one of these features appears
connected with the red giant L1421. \citet{G88} point out that
there is no compelling evidence of the connection of this high
temperature structure ($T=40^\circ~K$) with the cluster. However, the
detection of intrinsic polarization in L1421 (see Section 4), a likely
cluster member, suggests that this star is illuminating the
elongated cloud. Furthermore, the spatial orientation of this tail-like
 structure is well aligned with the direction of the spatial
proper motion of the cluster, corrected for the peculiar motion
of the sun and galactic rotation, based on the proper motion
study by \citet{CUD}, see Table 3 in \citep{KG95}.

In contrast with the bright non-variable red giants, almost all the
observed variable stars show significant polarization excesses, a result
consistent with those obtained by other authors and in the same direction
of arguments already given by \citet{FE88} in the sense that dust
formation and mass loss driven by stellar pulsations are connected.
In any case, and according to the very low mass loss rates obtained by
\citet{RAJO01}, who also point out the lack of dust enshrouded stars
in 47 Tuc, the variable stars do not appear as an important source
for dust injection (and replenishment) into the cluster environment.

In turn, the multicolor polarization observations of the cluster nucleus
do not show a detectable polarization maximum within the $\ubvr I$
wavelength range and, instead, display a smooth increase towards the $U$
band. This trend cannot be matched with a Serkowski law, even if we assume
rather extreme assumptions about $\lambda_\mathrm{max}$ and of the $K$
parameters that would imply values of the total to selective extinction ratio as
low as $R=2.0$ or less.  Besides, the fact that the $U$ polarization
exhibits significant changes with diaphragm aperture, and also a different
polarization angle, rather suggest that the observed behavior could be a
combination of the foreground interstellar polarization and of other
polarizing source that becomes more evident towards short wavelengths. The
blue post-asymptotic star known as BS or UIT7 may give an adequate
explanation if this object is in fact illuminating some dust located in
the cluster nucleus. Other possible origin, such as intrinsic polarization
associated with the atmospheres of giant stars within the 47\,Tuc nuclear
region can be ruled out as their ultraviolet magnitudes, around $U$=16 to
16.5, are some eight magnitudes fainter than the integrated brightness of
the nucleus within the $33 \arcsec$ diaphragm.

The detection of dust within the nuclear region of another metal rich
globular, NGC 6356, has been reported by \citet{HOPWOOD} on the basis
of observations carried out with $850 ~\micron$ arrays. However, ISO
observations of 47 Tuc itself \citep{ISO}, set a very low upper
limit for the dust mass (assuming silicate or amorphous carbon grains).
In fact, the peak emission at $120~\micron$ is offset from the cluster
center by $\sim 165\arcsec$ and a position angle of $118^\circ$ (see
figure 2 in that paper). This position angle is bracketed by those
corresponding to the bright-star-free regions $Z12$ and $Z13$ (although
they are closer to the cluster center). An analysis of the integrated
brightness of these regions, listed in Table~\ref{Table 4}, as a function
of position angle, shows that there is an overall decrease of the
cluster surface brightness in that sector (including Z14 and Z1) that,
 in the average, are 0.5 mag. fainter than the mean defined by
 all the regions. This obscuration
may arise in the same dust that reaches a flux of $\sim~35$~ MJy sr$^{-1}$
in the ISO map. The fact that this is probably a small angular scale
structure is suggested by the lack of a similar feature in the lower
resolution IRAS maps.

The origin of the disagreement between the ISO results, that suggest
that the nucleus is practically free of dust, and those coming from
the polarimetric results presented in this paper, remains
unclear. A preliminary hypothesis assumes that the
particles that produce polarization (in the cluster nucleus) and
those that may give rise to the $120~\micron$ emission (at the offset
 position) have markedly
different physical properties. The environmental conditions of
these particles may indeed be very different in terms, for example,
of the ionized gas density \citep{FREIRE} and its effect on the
grain survavility.

 High precision
color-magnitude diagrams (e.g., based on multicolor HST images) of
individual stars may provide a good test about the presence of dust
although, up to now, the published photometric results are still somewhat
noisy to reach definite conclusions. A dust cloud optical depth $\tau_U=0.5$
would imply a maximum blue extinction $A_B$ ranging from 0.4 to 0.2
magnitudes and a color excess $E(U-V)$ between 0.18 and 0.41 (for
a $\lambda^{-1}$ or a $\lambda^{-4}$ wavelength extinction dependence,
respectively). These values are compatible with the red envelope of
the $B$~vs.~$(U-V)$ color magnitude diagram presented by \citet{HGG00}
for stars within $21.54\arcsec$ from the cluster center.

A multicolor photometric program is currently under way
\citep{FF02}, and focused on a number of dark patches that might be
identified as knotty dust clouds with sizes smaller than 0.1 parsecs
inside a more diffuse dusty environment in the 47\,Tuc nuclear region.

\acknowledgements

The authors acknowledge the work of the CASLEO staff in developing
the polarimeter and also their enthusiastic help
during the several observing runs invested in this paper.
Useful remarks by an anonymous referee are appreciated.
This work was partially funded by CONICET and the Agencia
Nacional de Promoci\'on Cient\'{\i}fica y T\'ecnica de la
Rep.\ Argentina.

%\newpage

%\clearpage

\input{table1_pp.tex}
\clearpage
\input{table2_pp.tex}
\input{table3_pp.tex}
\input{table4_pp.tex}
\input{table5_pp.tex}

%\Table 1: Observed polarizations for the (non variable) stars.
%\check the case of BS. Columns: Number, alpha,delta, V, P,error,
%\theta, V, (B-V) (Chun-Freeman), VR (Mayor), Number Tucholke, p (tucholke)

%\Table 2 :Observed polarizations for stars between 4 and 20 arcmin.
%\ids from Lee (but watch out: some do not have Lee numbers...)

%\Table 3: Observed Polarizations for Variable Stars. Numbers from Sawyer.

%\Table 4: Observed polarizations for the star-free-regions.
%\Number, alpha, delta, V (aprox), P, error, theta, r(arcsecs), P.A.

%\Table 5: UBVRI polarization observations for the 47\,Tuc Nucleus.
%\Three values for U (17,33,45 arcsecs). Filter, P, error, theta, error.

\end{document}

%% file: table1_pp.tex
%\documentclass[preprint]{aastex}
%\begin{document}

\begin{deluxetable}{rccccccrlccr}
\tablecolumns{12}
\tablewidth{0pt}
\tablecaption{Visual Polarizations for Bright Giant Stars in 47\,Tuc\label{Table 1}}

\tablehead{

\colhead{N} & \colhead{$\alpha$\,{\small(J2000)}} &
 \colhead{$\delta$\,{\small(J2000)}} & \colhead{$V_{17\arcsec}$} &
 \colhead{$P_V$} &
% \colhead{$\epsilon_{P_V}$} &
 \colhead{$\theta_V$} & \colhead{T\tablenotemark{a}} &
 \colhead{mp\tablenotemark{b}} & \colhead{CF\tablenotemark{c}} &
 \colhead{$V$} & \colhead{$\bv$} & \colhead{R.V.} \\

\colhead{} & \colhead{hs\,:\,min\,:\,s} & \colhead{$^\circ\,:\,'\,:\,''$} &
 \colhead{mag} &
% \colhead{\%} &
 \colhead{\%} & \colhead{$^\circ$} & \colhead{} & \colhead{} &
 \colhead{} & \colhead{mag} & \colhead{mag} & \colhead{km/s}
 }

\startdata
1 & 00:23:29.4 & $-72$:06:20 & 11.27 & $0.47 \pm 0.05$ & 128.6 & 1175 &
98.3 & E175 & 11.63 & 1.86 & $-4.4$ \\

2 & 00:23:30.9 & $-72$:02:57 & 11.64 & $0.37 \pm 0.06$ & 120.8 & 1191 &
96.9 & E275 & 12.06 & 1.48 & $-27.9$ \\

3 & 00:23:34.1 & $-72$:03:29 & 12.02 & $0.31 \pm 0.06$ & 125.5 &
1225\tablenotemark{d} & 96.5 & E270 & 12.87 & 1.27 & $-35.8$ \\

4 & 00:23:35.0 & $-72$:07:39 & 11.43 & $0.30 \pm 0.05$ & 125.2
&\nodata&\nodata& E97 & 11.73 & 1.74 & $-10.1$ \\

5 & 00:23:35.3 & $-72$:01:30 & 11.67 & $0.41 \pm 0.06$ & 121.1 & 1243 &
23.7 & E323 & 12.12 & 1.51 & $-11.5$ \\

6 & 00:23:40.9 & $-72$:04:05 & 10.62 & $0.43 \pm 0.04$ & 125.2
&\nodata&\nodata& - &\nodata&\nodata&\nodata\\

7 & 00:23:41.6 & $-72$:06:45 & 11.16 & $0.35 \pm 0.05$ & 122.8 & 1320
& 0\phd\phn & E90 & 12.32 & 1.40 & $-10.3$ \\

8 & 00:23:44.8 & $-72$:05:34 & 11.61 & $0.41 \pm 0.06$ & 132.6
&\nodata&\nodata& F111 & 12.22 & 1.37 & $+1.6$ \\

9 & 00:23:47.1 & $-72$:05:53 & 10.99 & $0.34 \pm 0.05$ & 121.3
&\nodata&\nodata& F86 & 12.03 & 1.49 & $-6.0$ \\

10 & 00:23:47.3 & $-72$:06:53 & 11.31 & $0.31 \pm 0.05$ & 130.0
&\nodata&\nodata & E83 & 12.08 & 1.58 & $-22.3$ \\

11 & 00:23:47.7 & $-72$:02:49 & 11.15 & $0.36 \pm 0.05$ & 123.1 &
1379 & 0\phd\phn & E304 & 11.69 & 1.82 & $-27.8$ \\

12 & 00:23:49.0 & $-72$:03:13 & 11.42 & $0.30 \pm 0.06$ & 123.5
&\nodata&\nodata & F218 & 12.13 & 1.49 & $-13.0$ \\

13 & 00:23:49.3 & $-72$:06:20 & 11.24 & $0.41 \pm 0.05$ & 130.4 &
1390 & 98.3 & F66 & 12.00 & 1.50 & $-8.3$ \\

14 & 00:23:49.5 & $-72$:05:30 & 11.05 & $0.38 \pm 0.05$ & 120.0
&\nodata&\nodata & F106 & 12.07 & 1.50 & $-5.6$ \\

15 & 00:23:50.4 & $-72$:05:51 & 11.16 & $0.38 \pm 0.06$ & 130.0
&\nodata&\nodata & F89 & 12.21 & 1.69 & $-13.5$ \\

16 & 00:23:50.5 & $-72$:04:21 & 10.45 & $0.38 \pm 0.04$ & 124.6
&\nodata&\nodata & F188 & 11.48 & 2.28 & $-26.8$ \\

17 & 00:23:51.2 & $-72$:03:49 & 10.82 & $0.44 \pm 0.04$ & 121.4
&\nodata&\nodata & F203 & 11.70 & 1.77 & $-7.7$ \\

18 & 00:23:53.1 & $-72$:05:07 & 10.11 & $0.34 \pm 0.03$ & 121.0
&\nodata&\nodata &\nodata&\nodata&\nodata&\nodata\\

19 & 00:23:53.1 & $-72$:04:15 & 10.56 & $0.35 \pm 0.04$ & 120.6
&\nodata&\nodata & F193 & 11.62 & 1.86 & $-29.1$ \\

20 & 00:23:54.5 & $-72$:06:38 & 11.29 & $0.39 \pm 0.05$ & 116.4
&\nodata&\nodata & F60 & 12.34 & 1.37 & $-14.0$ \\

21 & 00:23:54.6 & $-72$:03:39 & 10.94 & $0.48 \pm 0.05$ & 123.3
&\nodata&\nodata & F229 & 11.71 & 1.76 & $-31.4$ \\

22 & 00:23:54.8 & $-72$:05:08 & 10.97 & $0.36 \pm 0.05$ & 123.8
&\nodata&\nodata &\nodata&\nodata&\nodata&\nodata\\

23 & 00:23:56.0 & $-72$:01:48 & 11.57 & $0.43 \pm 0.06$ & 121.1
&\nodata&\nodata & E354 & 12.44 & 1.35 & $-14.9$ \\

24 & 00:23:58.0 & $-72$:05:49 & 10.64 & $0.39 \pm 0.05$ & 122.0
&\nodata&\nodata &\nodata&\nodata&\nodata&\nodata\\

25 & 00:23:58.1 & $-72$:03:31 & 10.97 & $0.35 \pm 0.05$ & 122.2
&\nodata&\nodata & F260 & 11.74 & 1.61 & $-24.4$ \\

26 & 00:23:59.0 & $-72$:02:35 & 11.27 & $0.39 \pm 0.05$ & 124.3 & 1480 &
98.3 & E370 & 11.64 & 1.94 & 10.3 \\

27 & 00:24:01.5 & $-72$:03:01 & 11.36 & $0.38 \pm 0.05$ & 125.0
&\nodata&\nodata & F276 & 12.03 & 1.50 & $-29.9$ \\

28 & 00:24:02.9 & $-72$:08:11 & 11.66 & $0.34 \pm 0.06$ & 125.1
&\nodata&\nodata & E44 & 12.20 & 1.43 & $-14.4$ \\

29 & 00:24:04.2 & $-72$:06:55 & 10.95 & $0.36 \pm 0.05$ & 127.4
&\nodata&\nodata & F1 & 11.90 & 1.51 & $-25.5$ \\

30 & 00:24:06.5 & $-72$:07:00 & 12.30 & $0.24 \pm 0.06$ & 124.2
&\nodata&\nodata & F588 & 12.34 & 1.37 & $-22.5$ \\

31 & 00:24:07.1 & $-72$:03:30 & 10.71 & $0.39 \pm 0.05$ & 124.3
&\nodata&\nodata & F290 & 11.86 & 1.62 & $-26.7$ \\

32 & 00:24:08.8 & $-72$:02:59 & 10.95 & $0.33 \pm 0.05$ & 122.5
&\nodata&\nodata & F305 & 11.53 & 1.95 & $-25.3$ \\

33 & 00:24:10.8 & $-72$:07:01 & 12.00 & $0.29 \pm 0.06$ & 125.5
&\nodata&\nodata & F563 & 12.60 & 1.28 &\nodata\\

34 & 00:24:13.0 & $-72$:06:50 & 12.25 & $0.32 \pm 0.06$ & 140.5
&\nodata&\nodata & F561 & 12.87 & 1.15 &\nodata\\

35 & 00:24:14.3 & $-72$:06:35 & 11.94 & $0.32 \pm 0.06$ & 124.4
&\nodata&\nodata & F553 & 12.19 & 1.29 & $-9.2$ \\

36 & 00:24:14.8 & $-72$:04:45 & 10.50 & $0.38 \pm 0.04$ & 127.3
&\nodata&\nodata &\nodata&\nodata&\nodata&\nodata\\

37 & 00:24:16.2 & $-72$:01:32 & 11.45 & $0.37 \pm 0.06$ & 122.1
&\nodata&\nodata & E397 & 12 & 1.91 & $-20.3$ \\

38 & 00:24:18.5 & $-72$:04:44 & 10.11 & $0.41 \pm 0.03$ & 126.4
&\nodata&\nodata &\nodata&\nodata&\nodata&\nodata\\

39 & 00:24:19.3 & $-72$:05:19 & 11.20 & $0.48 \pm 0.05$ & 117.2
&\nodata&\nodata & F478 & 11.93 & 1.58 & $-36.2$ \\

40 & 00:24:20.3 & $-72$:05:52 & 11.08 & $0.42 \pm 0.05$ & 122.3
&\nodata&\nodata & F490 & 12.08 & 1.44 & $-19.8$ \\

41 & 00:24:20.5 & $-72$:05:59 & 11.63 & $0.37 \pm 0.06$ & 118.6
&\nodata&\nodata & F507 & 12.17 & 1.42 & $-9.6$ \\

42 & 00:24:21.0 & $-72$:03:33 & 11.16 & $0.37 \pm 0.05$ & 125.2
&\nodata&\nodata & F338 & 11.80 & 1.70 &\nodata\\

43 & 00:24:23.4 & $-72$:04:06 & 10.92 & $0.36 \pm 0.05$ & 122.1 &
1700 & 98.4 &\nodata&\nodata&\nodata&\nodata\\

44 & 00:24:24.8 & $-72$:04:49 & 10.87 & $0.37 \pm 0.05$ & 117.5
&\nodata&\nodata & F429 & 11.80 & 1.68 & $-24.5$ \\

45 & 00:24:30.0 & $-72$:04:32 & 11.55 & $0.42 \pm 0.06$ & 121.1 &
1774 & 56.6 & F400 & 11.82 & 1.69 & $-27.8$ \\

46 & 00:24:30.7 & $-72$:07:05 & 12.16 & $0.39 \pm 0.06$ & 126.7
&\nodata&\nodata & E630 & 12.88 & 1.20 &\nodata\\

47 & 00:24:30.9 & $-72$:07:31 & 12.65 & $0.25 \pm 0.07$ & 127.2
&\nodata&\nodata & E629 & 12.86 & 1.10 &\nodata\\

48 & 00:24:33.9 & $-72$:06:18 & 11.92 & $0.34 \pm 0.06$ & 120.2 & 1816
& 15.4 & E587 & 12.77 & 1.27 & $-12.2$ \\

49 & 00:24:34.1 & $-72$:05:13 & 12.07 & $0.43 \pm 0.06$ & 120.5
&\nodata&\nodata & F445 & 13.05 & 1.20 &\nodata\\

50 & 00:24:36.5 & $-72$:05:40 & 11.34 & $0.43 \pm 0.05$ & 123.2
&\nodata&\nodata & E582 & 11.95 & 1.62 & $-26.5$ \\

51 & 00:24:37.6 & $-72$:05:11 & 11.73 & $0.31 \pm 0.06$ & 122.6 &
1862 & 0\phd\phn & E647 & 12.78 & 1.10 &\nodata\\

52 & 00:24:41.6 & $-72$:04:31 & 11.76 & $0.36 \pm 0.06$ & 122.8 &
1911 & 97.6 & E512 & 12.62 & 1.27 & $-19.4$ \\

53 & 00:24:43.8 & $-72$:07:23 & 11.62 & $0.43 \pm 0.06$ & 127.7 &
1940 & 96.4 & E620 & 12.00 & 1.59 & $-24.3$ \\
\enddata

\tablenotetext{a}{Identification number from Tucholke 1992.}
\tablenotetext{b}{Membership probability from Tucholke 1992.}
\tablenotetext{c}{Identification number from Chun \& Freeman 1978.}
\tablenotetext{d}{Also listed as star 1222, with mp=0.}
\end{deluxetable}
%\end{document}

%% file: table2_pp.tex
%\documentclass[preprint]{aastex}
%\begin{document}

\begin{deluxetable}{crrcc}
\tablecolumns{5}
\tablewidth{0pc}
\tablenum{2}
\tablecaption{Visual Polarizations for Stars Between 4 and 20 arcmin from
 the 47\,Tuc Nucleus
\label{Table 2}}
\tablehead{
\colhead{L\tablenotemark{a}
} & \colhead{X} & \colhead{Y} & \colhead{$P_V$}
&%\colhead{$\epsilon_{P_V}$} &
 \colhead{$\theta_V$} \\

\colhead{} & \colhead{$''$} & \colhead{$''$} & \colhead{\%} &
%\colhead{\%} &
 \colhead{$^\circ$}
}

\startdata
 1421 & $-190.0$ & $-1060.0$ & $0.53 \pm 0.05$ & 101.4 \\
 2620 & $-555.9$ &  $-167.2$ & $0.43 \pm 0.05$ & 123.7 \\
 2705 & $-492.0$ &   $-73.8$ & $0.50 \pm 0.07$ & 132.3 \\
 2758 & $-423.1$ &  $-373.9$ & $0.42 \pm 0.04$ & 112.5 \\
 3708 & $-546.1$ &     64.0  & $0.39 \pm 0.05$ & 110.4 \\
 3730 & $-418.2$ &    167.4  & $0.44 \pm 0.05$ & 121.4 \\
 3736 & $-433.0$ &    196.9  & $0.44 \pm 0.06$ & 118.1 \\
 4715 & $-275.5$ &    260.8  & $0.41 \pm 0.04$ & 121.3 \\
 4728 &    50.0  &    541.3  & $0.48 \pm 0.05$ & 111.5 \\
 4729 &    50.0  &    492.1  & $0.37 \pm 0.06$ & 117.9 \\
 5312 &   806.9  &    885.7  & $0.47 \pm 0.06$ & 118.0 \\
 5529 &   561.0  &    659.4  & $0.46 \pm 0.05$ & 114.7 \\
 5622 &   226.3  &    551.1  & $0.40 \pm 0.05$ & 104.9 \\
 8756 &   255.8  &  $-295.2$ & $0.28 \pm 0.07$ & 121.6 \\
\enddata 
\tablenotetext{a}{Identification number from Lee 1977}
\end{deluxetable}
%\end{document}

%% file: table3_pp.tex
%\documentclass[preprint]{aastex}
%\begin{document}

\begin{deluxetable}{lcrrr}
\tablecolumns{5}
\tablewidth{0pc}
\tablenum{3}
\tablecaption{Visual Polarizations for Variable Stars in 47\,Tuc
\label{Table 3}}
\tablehead{
\colhead{Star\tablenotemark{a}} & \colhead{$P_V$} &
%\colhead{$\epsilon_{P_V}$} &
 \colhead{$\theta_V$} & \colhead{$U$} &
\colhead{$Q$}\\

\colhead{} & \colhead{\%} &% \colhead{\%} &
 \colhead{\%} & \colhead{\%} & \colhead{\%}
}

\startdata
V1 & $0.36 \pm 0.06$ & 99.0 & $-$0.11 & $-$0.34 \\
V2 & $0.06 \pm 0.05$ & 45.0 & 0.06 & 0.00 \\
V3 & $0.42 \pm 0.05$ & 105.1 & $-$0.21 & $-$0.36 \\
V4 & $0.47 \pm 0.08$ & 130.7 & $-$0.46 & $-$0.07 \\
V4 & $0.23 \pm 0.04$ & 52.6 & 0.22 & $-$0.06 \\
V6 & $0.36 \pm 0.06$ & 135.8 & $-$0.36 & 0.01 \\
V8 & $0.47 \pm 0.04$ & 1.8 & 0.03 & 0.47 \\
V10 & $0.31 \pm 0.04$ & 120.5 & $-$0.27 & $-$0.15 \\
V11 & $0.47 \pm 0.06$ & 135.6 & $-$0.47 & 0.01 \\
V13 & $2.37 \pm 0.10$ & 123.9 & $-$2.20 & $-$0.90 \\
\enddata 
\tablenotetext{a}{Identification from Sawyer-Hogg 1973}
\end{deluxetable}
%\end{document}

%% file: table4_pp.tex
%\documentclass[preprint]{aastex}
%\begin{document}

\begin{deluxetable}{rcccccr}
\tablecolumns{7}
\tablewidth{0pc}
\tablenum{4}
\tablecaption{Visual Polarization for the Bright-Star-Free Regions in
the Field of 47\,Tuc
\label{Table 4}}
\tablehead{

\colhead{N} & \colhead{R.A.\,{\small(J2000)}} &
 \colhead{Dec\,{\small(J2000)}} &
 \colhead{$V_{17\arcsec}$} & \colhead{$P_V$} &% \colhead{$\epsilon_{P_V}$} &
 \colhead{$\theta_V$} & \colhead{P.A.}\\

\colhead{} & \colhead{hs\,:\,min\,:\,s} & \colhead{$^\circ\,:\,'\,:\,''$} &
 \colhead{mag} & \colhead{\%} &% \colhead{\%} &
 \colhead{$^\circ$} & \colhead{$^\circ$}
}

\startdata
% N    R.A. (J2000)   Dec(J2000)    V(17)   Pv     Epv  Theta   P.A.
1 & 00:24:01.84 & $-72$:07:01.8 & 12.78 & $0.35 \pm 0.06$ & 119.5 & 188.4 \\
2 & 00:23:52.40 & $-72$:06:31.0 & 11.78 & $0.36 \pm 0.04$ & 123.5 & 212.5 \\
3 & 00:23:43.94 & $-72$:05:41.3 & 12.23 & $0.38 \pm 0.05$ & 120.7 & 244.6 \\
4 & 00:23:44.13 & $-72$:05:11.8 & 11.89 & $0.35 \pm 0.04$ & 125.0 & 259.3 \\
5 & 00:23:44.23 & $-72$:04:23.8 & 11.67 & $0.35 \pm 0.04$ & 125.8 & 286.1 \\
6 & 00:23:49.56 & $-72$:03:52.4 & 12.03 & $0.36 \pm 0.04$ & 126.0 & 308.6 \\
7 & 00:23:55.85 & $-72$:03:23.4 & 11.99 & $0.43 \pm 0.04$ & 123.2 & 331.2 \\
8 & 00:24:06.92 & $-72$:03:15.0 & 11.83 & $0.37 \pm 0.03$ & 122.1 & 2.6 \\
9 & 00:24:18.85 & $-72$:03:16.7 & 11.67 & $0.35 \pm 0.04$ & 124.2 & 31.8 \\
10 & 00:24:17.14 & $-72$:03:40.6 & 11.86 & $0.37 \pm 0.04$ & 124.8 & 53.5 \\
11 & 00:24:30.56 & $-72$:04:43.9 & 11.49 & $0.39 \pm 0.03$ & 122.0 & 85.4 \\
12 & 00:24:27.96 & $-72$:05:38.0 & 12.72 & $0.40 \pm 0.06$ & 122.4 & 113.8 \\
13 & 00:24:22.37 & $-72$:06:10.7 & 12.39 & $0.39 \pm 0.05$ & 126.7 & 135.8 \\
14 & 00:24:16.61 & $-72$:06:46.5 & 12.41 & $0.37 \pm 0.06$ & 120.3 & 156.5 \\
\enddata
 
\end{deluxetable}
%\end{document}

%% file: table5_pp.tex
%\documentclass[preprint]{aastex}
%\begin{document}

\begin{deluxetable}{lccccc}
\tablecolumns{6}
\tablewidth{0pc}
\tablenum{5}
\tablecaption{$UBVRI$ Polarizations for the Nucleus of 47\,Tuc
\label{Table 5}}
\tablehead{

\colhead{Filter} & \colhead{$\lambda$} & \colhead{$P_\lambda$} &
%\colhead{$\epsilon_{P_\lambda}$} &
 \colhead{$\theta_\lambda$} &
%\colhead{$\epsilon_{\theta_\lambda}$}
 \\

\colhead{} & \colhead{nm} & \colhead{\%} &% \colhead{\%} &
 \colhead{$^\circ$} & % \colhead{$^\circ$}
}

\startdata
$U$ ($17\arcsec$) & 360 & $0.44 \pm 0.11$ & $127.4 \pm 2.5$ \\
$U$ ($33\arcsec$) & 360 & $0.72 \pm 0.06$ & $118.6 \pm 2.5$ \\
$U$ ($45\arcsec$) & 360 & $0.48 \pm 0.06$ & $116.8 \pm 2.0$ \\
$B$ & 430 & $0.46 \pm 0.02$ & $125.6 \pm 1.0$ \\
$V$ & 550 & $0.36 \pm 0.01$ & $124.5 \pm 0.5$ \\
$R$ & 630 & $0.33 \pm 0.01$ & $124.4 \pm 0.5$ \\
$I$ & 780 & $0.29 \pm 0.03$ & $123.8 \pm 1.5$ \\
\enddata
 
\end{deluxetable}
%\end{document}

%% file: Forte_pp.bbl
\begin{thebibliography}

\bibitem[Auriere \& Leroy(1990)]{AL90}Auriere, M., \& Leroy, J. L. 1990,
\aap, 234, 164

\bibitem[Chun \& Freeman(1978)]{CF78}Chun, M. S., \& Freeman, K. C. 1978,
\aj, 83, 376

\bibitem[Cramer(1984)]{CRA84}Cramer, N. 1984, \aap, 132, 283

\bibitem[Clarke \& Stewart(1986)]{CLST}Clarke, D. \& Stewart, B.G.
Vistas Astron., 29, 27.

\bibitem[Cudworth \& Hanson(1993)]{CUD}Cudworth, K. M., \& Hanson,
R. B. 1993, \aj, 105, 168

\bibitem[Dixon et al.(1995)Dixon, Davidsen, \& Ferguson]{DI95}Dixon,
W. V. D., Davidsen, A. F., \& Ferguson, H. C. 1995, \apj, 454, L47

\bibitem[Draine \& Lee (1984)]{DRL}Draine, B.T., \& Lee, H.M. 1984, \apjs,
285, 89

\bibitem[Faifer \& Forte(2002)]{FF02}Faifer, F., \& Forte, J. C. 2002,
in prep.

\bibitem[Flower(1996)]{FL96}Flower, P. J. 1996, \apj, 469, 355

\bibitem[Forte \& M\'endez(1988)]{FM88}Forte, J. C., \& M\'endez,
 M. 1988, \aj, 95, 500

\bibitem[Forte \& M\'endez(1989)]{FM89}Forte, J. C., \& M\'endez,
 M. 1989, \apj, 345, 222

\bibitem[Freire et al. (2001)]{FREIRE}Freire, P. C., Kramer, M.,
Lyne, A. G., Camilo, F., Manchester, R. N., \& D'amico, N. D. 2001,
\apj, 557, 105

\bibitem[Frogel \& Elias(1988)]{FE88}Frogel, J. A., \& Elias, J. H.
1988, \apj, 324, 823

\bibitem[Gillett et al.(1988)]{G88}Gillett, F. C., de Jong, T., Neugebauer, G.,
Rice, W. L., \& Emerson, J. P. 1988 \aj, 96, 116

\bibitem[Guhathakurta et al.(1992)] {GYSB}Guhathakurta, P., Yanny, B.,
Schneider, D. P., \& Bahcall, J. N. 1992, \aj, 104, 1790

\bibitem[Hesser et al.(1987)]{HESSER}Hesser, J. E., Harris, W. E.,
VandenBerg, D. A., Allwright, J. W. B., Shott, P., \& Stetson,
P. B. 1987, \pasp, 99, 739

\bibitem[Hopwood et al.(1998)]{HOPWOOD}Hopwood, M. E. L., Evans, A.,
Penny, A., Eyres, S. P. S. 1998, \mnras, 301, L30

\bibitem[Hopwood et al.(1999)]{ISO}Hopwood, M. E. L., Eyres, S. P. S.,
Evans, A., Penny, A. \& Odenkirchen, M. 1999, \aap, 350, 49

\bibitem[Howell et al.(2000)Howell, Guhathakurta, \& Gilliland]
{HGG00} Howell, J. H., Guhathakurta, P., \& Gilliland, R. L. 2000,
\pasp, 112, 1200

\bibitem[Jakobsen et al.(1987)Jakobsen, de Vries, \& Paresce]{JAKOB}
Jakobsen, P., de Vries, J. S., \& Paresce, F. 1987, \aap, 183, 335

\bibitem[Jura(1978)]{JU78}Jura, M. 1978, \apj, 223, 421

\bibitem[Kanagy \& Wyatt(1978)]{KW78}Kanagy, S. P., \& Wyatt,
S. P. 1978, \aj, 83, 779

\bibitem[Knapp et al.(1995)Knapp, Gunn, \& Connolly]{K95}Knapp, G. R.,
Gunn, J. E., \& Connolly, A. J. 1995, \apj, 448, 195

\bibitem[Krockenberger \& Grindlay(1995)]{KG95}Krockenberger, M., \& Grindlay,
J. E. 1995, \apj, 451 200

\bibitem[Lee(1977)]{LEE77}Lee, S.-W. 1977, \aap, 27, 381

\bibitem[Lloyd Evans(1974)]{LLOYD}Lloyd Evans, T. 1974, \mnras, 167,
393

\bibitem[Magalh\~aes et al.(1984)Magalh\~aes, Benedetti, \& Roland]{MA84}
Magalh\~aes, A. M., Benedetti, E., \& Roland, E. H. 1984, \pasp, 96, 383

\bibitem[Martin \& Shawl(1981)]{MS81}Martin, P. G., \& Shawl, S. J. 1981,
\apj, 251, 108

\bibitem[Mart\'{\i}nez et al.(1990)]{CASPROF}Mart\'{\i}nez, E.,
Aballay, J. L., Mar\'un, A., \& Ruar\-tes, H. 1990, Bol.\ Asoc.\ Arg.\
de Astronom\'{\i}a, 36, 342

\bibitem[Mathewson \& Ford(1970)]{MF}Mathewson, D. S., \& Ford, V. L. 1970,
 \memras, 74, 139

\bibitem[Mayor et al.(1983)]{MAYOR}Mayor, M., et al. 1983, \aaps, 54, 495

\bibitem[Mayor et al.(1984)]{MAYOR2}Mayor, M., et al. 1984, \aap, 134, 118

\bibitem[Minniti et al.(1992)Minniti, Coyne \& Clari\'a]{MI92}
Minniti, D., Coyne, G. V. \& Clari\'a, J. J. 1992, \aj, 103, 871

\bibitem[Minniti et al.(1990)Minniti, Coyne, \& Tapia]{MI90}Minniti,
D., Coyne, G. V., \& Tapia, S.  1990, \aap, 236, 371

%\bibitem[O'Connell et al.(1997)]{OC97}O'Connell, R.W., Dorman, B., Sha, R.Y.,
%Rood, T.R., Landsman, W.B., Witt, A.N., Bohlin, R.C., Neff, S.G., Roberts,
%M. S., Smith, A.M. \& Stecher, T.P. 1997, \aj 114, 1982

\bibitem[O'Connell et al.(1997)]{OC97}O'Connell, R. W., et al. 1997,
\aj,114, 1982

\bibitem[Origlia et al.(1997)]{OR97}Origlia, L., Scaltriti, F.,
Anderlucci, E., Ferraro, F. R., \& Fusi Pecci, F. 1997, \mnras, 292,
753

\bibitem[Pellizza Gonz\'alez et al.(2002)Pellizza Gonz\'alez, Forte,
\& Carpintero] {PELLIZZA} Pellizza, L., Forte, J. C., \& Carpintero,
D.  2002, in prep.

\bibitem[Ramdani \& Jorissen (2001)]{RAJO01}Ramdani, A., \& Jorissen, A. 2001,
 \aap,372, 85

\bibitem[Reed et al.(1988)Reed, Hesser, \& Shawl]{REED}Reed, B. C.,
Hesser, J. E., \& Shawl, S. J.  1988, \pasp, 100, 545

\bibitem[Roberts(1960)]{ROB}Roberts, M. S. 1960, \aj, 65, 457

\bibitem[Sawyer-Hogg(1973)]{SAWYER}Sawyer-Hogg, H. 1973, Publ.\ DDO 3, No 6

\bibitem[Schlegel et al.(1999)Schlegel, Finkbeiner, \& Davis]{SCHLE}
Schlegel, D. J., Finkbeiner, D. P., \& Davis, M. 1998 \apj, 500, 525

\bibitem[Sellgren et al.(1983)Sellgren, Werner, \& Dinerstein]{SELLGREN}
Sellgren, K., Werner, M. W., \& Dinerstein, H. L. 1983, \apj, 271, L13

\bibitem[Serkowski(1973)]{SK73}Serkowski, K. 1973, in Proc.\ IAU
Symposium No 52, Interstellar Dust and Related Topics, eds.\
J. M. Greenberg, \& H. C. van de Hulst (Dordrecht: Reidel), 145

\bibitem[Serkowski et al.(1975)Serkowski, Mathewson, \& Ford]{SMF75}
Serkowski, K., Mathewson, D. S., \& Ford, V. L. 1975, \apj, 196, 261

\bibitem[Tucholke(1992)]{TU92}Tucholke, H. J. 1992, \aaps, 93, 293

\bibitem[Van de Hulst(1967)]{VdH}van de Hulst, H. C. 1967, in Light Scattering
by Small Particles, Dover Inc. 1981
%\bibitem[Murthy et al. 1992]{MURTHY}Murthy, J., Henry, R.C. \& Holberg, J. B.
%\1994, \apj, 428, 233

\bibitem[Whittet et al.(1992)]{WHIT92}Whittet, D. C. B., Martin,
P. G., Hough, J. H., Rouse, M. F., Bailey, J. A., \& Axon, D. J. 1992,
\apj, 386, 562

\bibitem[Wilking et al.(1982)Wilking, Lebofsky, \& Rieke]{WILL82}
Wilking, B. A., Lebofsky, M. J., \& Rieke, G. H. 1982, \aj, 87, 695

\bibitem[Witt et al.(1992)Witt, Thronson, \& Capuano]{WITT}Witt, A. N.,
Thronson, H. A., \& Capuano, J. M. 1992, \apj, 393, 611
\end{thebibliography}
